\documentclass[a4paper,11pt]{article}
\usepackage{jcappub} 
\usepackage{lineno}
\usepackage{hyperref}
\usepackage{rotating}
\usepackage{aas_macros}
\usepackage{booktabs}
\usepackage{tablefootnote}

\arxivnumber{} 

\title{Reducing nuisance prior sensitivity via non-linear reparameterization, with application to EFT analyses of large-scale structure}

\author[a,b,e]{S. Paradiso\note{Corresponding author.},}
\author[a,b]{M. Bonici,}
\author[c]{M. Chen,}
\author[a,b,d]{W.J. Percival,}
\author[f,g]{G. D'Amico,}
\author[a,b]{H. Zhang,}
\author[c]{G. McGee}

\affiliation[a]{Waterloo Centre for Astrophysics, University of Waterloo,
Waterloo, ON N2L 3G1, Canada}
\affiliation[b]{Department of Physics and Astronomy, University of Waterloo,
Waterloo, ON N2L 3G1, Canada}
\affiliation[c]{Department of Statistics and Actuarial Sciences, University of Waterloo,\\
Waterloo, ON N2L 3G1, Canada}
\affiliation[d]{Perimeter Institute for Theoretical Physics,
31 Caroline St North, Waterloo, ON N2L 2Y5, Canada}
\affiliation[e]{INAF, Istituto di Astrofisica Spaziale e Fisica Cosmica di Bologna,\\via P. Gobetti 101, I-40129 Bologna, Italy}
\affiliation[f]{Department of Mathematical, Physical and Computer Sciences, University of Parma,\\ Parco Area delle Scienze 7/a, 43124 Parma, Italy}
\affiliation[g]{INFN Gruppo Collegato di Parma,\\Parco Area delle Scienze 7/a, 43124 Parma, Italy}

\emailAdd{simone.paradiso@uwaterloo.ca}

\abstract{Many physical models contain nuisance parameters that quantify unknown properties of an experiment that are not of primary relevance. Typically, these cannot be measured except by fitting the models to the data from the experiment, requiring simultaneous measurement of interesting parameters that are our target of inference and nuisance terms that are not directly of interest. A recent example of this is fitting Effective Field Theory (EFT) models to large-scale structure (LSS) data to make cosmological inferences. These models have a large number of nuisance parameters that are typically correlated with cosmological parameters in the posterior, leading to strong dependence on the nuisance parameter priors. We introduce a reparametrization method that leverages Generalized Additive Models (GAMs) to decorrelate nuisance parameters from the parameters of interest in the likelihood, even in the presence of non-linear relationships. This reparametrization forms a natural basis within which to define priors that are independent between nuisance and target parameters: the separation means that the marginal posterior for cosmological parameters does not depend on simple priors placed on nuisance terms. 
In application to EFT models using LSS data, we demonstrate that the proposed approach leads to robust cosmological inference.}

\begin{document}
\maketitle
\flushbottom

\section{Introduction}
\label{sec:Intro}

Measuring cosmological parameters from observations of the Universe is a fundamental goal of modern cosmology. The properties of the Universe depend on many different physical processes, enabling a multitude of different measurements \cite{lss_review,Planck2018-VI}. In general, these physical processes are well understood on large-scales, where perturbations are small and the linear limit is valid for most of the physics (e.g. \cite{Lewsi2000,Kaiser1987}). However, on smaller scales, this endeavour is complicated by unresolved small-scale physics \cite{Bernardeau2002,Desjacques2018} and systematic uncertainties in observational data \cite{Bianchi2024}. The dependence on the cosmological parameters eventually becomes so weak and complicated on small-scales that it is not straightforward to link observation and cosmological properties except through careful forward modelling \cite{Jasche2013}. Even with forward modelling, one must include all of the physical and observational effects, and be confident in their modelling. Given these difficulties, it is common to account for the unknown effects on quasi-linear scales and the mildly non-linear regime using nuisance parameters controlling likely deviations in models (for example, for unknown baryonic effects \cite{Eifler2015}). The nuisance terms often affect the model in a similar way to the physics that we do understand, leading to correlations between nuisance parameters and cosmological parameters, and degeneracies between the two in Bayesian posteriors \cite{Trotta2008}.

Priors for nuisance parameters are often chosen somewhat arbitrarily, as inference for these parameters does not carry easy to access cosmological information, but degeneracies between these nuisance and cosmological parameters causes cosmological inferences to be sensitive to choice of nuisance priors. There can be dependencies on the size of the region allowed for the nuisance terms, and on the location of the priors, which can drive the final cosmological Bayesian inferences if they pull away from the values preferred by the data. These are called the prior volume and prior weight effects, respectively in \cite{DESI2024-fullshape}. In a Bayesian framework the choice of prior can  impact the posterior distribution, so using arbitrarily chosen priors for nuisance parameters can influence the inferred posterior distributions of cosmological parameters in undesirable ways, leading to biased or misleading results \cite{bayesian_nuisance_parameters}. 

Consider as motivating example the Effective Field Theory (EFT) framework for modelling large-scale structure, which has gained prominence due to its ability to systematically incorporate non-linear corrections from small-scale physics into cosmological models \cite{eft_nuisance_correlation,Porto2014TheStructures,Perko2016BiasedStructure}. EFT introduces a set of nuisance parameters that are essential for capturing the small-scale behaviour of the galaxy distribution but are degenerate with cosmological parameters. The model has been used to extract information from the Baryon Oscillation Spectroscopic Survey (BOSS \cite{Dawson2013TheSDSS-III}) by \cite{dAmico2020TheStructure,Ivanov2020CosmologicalSpectrum}, and its extension, eBOSS \cite{Dawson2016TheData} by \cite{eBOSS-QSO-EFT2,eBOSS-QSO-EFT1}. An investigation into the effect of nuisance parameter priors was completed by \cite{Simon:2022lde,Chudaykin:2024wlw}, while a comparison to frequentist results was presented in \cite{Holm2023BayesianData}. Most recently, an EFT-based analysis of the Dark Energy Spectroscopic Survey (DESI \cite{DESICollaboration2016TheDesign,DESICollaboration2016TheDesignb,Collaboration2022OverviewInstrument}) data was presented in \cite{DESI2024-fullshape}. Moreover, the bispectrum analysis presented in~\cite{DAmico:2022osl} showed that such projection effects vanish when a bigger volume is considered (\textit{i.e.} when the measurement error was smaller), as in that case the data are sufficiently constraining (see also~\cite{Hadzhiyska2023Cosmology}).

The results of these analyses are sensitive to choice of nuisance parameter priors, and several solutions have been proposed to address this issue. Some approaches involve imposing specific, physically motivated priors on nuisance parameters to constrain their effect on cosmological inference \cite{informative_priors_nuisance}. For instance, Halo Occupation Distribution (HOD) models often incorporate priors based on empirical or simulation-derived expectations of how galaxies populate dark matter halos, offering additional constraints that reduce the parameter space \cite{Zhang:2024, Zhang:2025yco, Ivanov:2024xgb}. Alternatively, choosing among so-called non-informative priors like uniform or Jeffreys priors present other complications. Uniform priors are often employed for their simplicity but can inadvertently impose subjective assumptions about the scale of the parameter space. Jeffreys priors, being invariant under re-parameterization \citep{Jeffreys1939}, are theoretically appealing, but in practice, they are difficult to compute when the model’s dependence on nuisance parameters is highly non-linear and non-additive \cite{Hadzhiyska2023Cosmology, McCann:2023}. While these methods offer some relief from the degeneracies between cosmological and nuisance parameters, they do not fully address the projection effects, and prior dependence remains an issue. Alternatively, frequentist methodologies, which rely on profile likelihood techniques, are less sensitive to the impact of nuisance parameters, matching them to their best-fit values rather than marginalizing over them \cite{frequentist_nuisance_solutions}. 

In this paper, we propose a methodological solution to address this challenge. Rather than proposing a particular prior, we instead propose a re-parameterization of the model, after which cosmological inference is robust to choice of independent nuisance  priors. This is similar in concept to the idea of carefully choosing combinations of parameters to be fitted in order to break degeneracies between bias and cosmology, as undertaken by \cite{Tucci_2024,Chen_2024}, but we adopt a more formal statistical method for the re-parameterization chosen.

Our proposed re-parameterization strategy leverages Generalized Additive Models (GAMs) to systematically remove the dependence of cosmological parameters on nuisance parameter priors \cite{gam_models}. After re-parametrization, the cosmological parameters are uncorrelated with the nuisance parameters in the likelihood, which becomes approximately separable, so any independent nuisance parameter prior will not affect the marginal posterior for cosmological parameters. Consequently, we eliminate much of the prior projection effect that can distort the inference process, leading to more robust cosmological parameter estimates. This approach allows us to dramatically reduce degeneracies, enhancing the accuracy of cosmological measurements without sacrificing precision \cite{gam_cosmo_precision}. And unlike  Jeffreys priors, the methodology is computationally straightforward to implement even in complex models.

The rest of this paper is structured in the following way: In Sec~\ref{sec:Methods} we introduce the methodological framework, describing the re-parameterization approach through a simple linear relation and GAM. In Sec.~\ref{sec:EFT} we briefly describe the EFT. We then describe the simulation pipeline  in Sec.~\ref{sec:dataset} and present our results in Sec.~\ref{sec:Results}, followed by conclusions in Sec.~\ref{sec:Conclusions}.

\section{Methods}
\label{sec:Methods}

We present a novel methodological approach to relieve the correlation between target cosmological parameters and nuisance parameters.

\subsection{Parameter Orthogonalization}
\label{subsec:least-square}
Let $\theta=[C^T,N^T]^T$ be a vector of parameters that can be partitioned into cosmological parameter $C$ and nuisance parameter $N$. We have useful prior information on $C$ but little sense of what prior to specify for nuisance parameters $N$. Although $N$ is not itself of interest, marginal inference about $C$ can be sensitive to choice of prior for $N$. This is because the marginal posterior of $C$ and $N$, $\pi(N,C|y)$ where $y$ represents the data, are not necessarily separable even if we adopt independent priors $\pi(N,C)=\pi(N)\pi(C)$ because of coupling through the likelihood:

 \begin{align*}
 \pi(N,C|y)\propto \pi(N)\pi(C)f(y| N,C) \neq \pi(N|y)\pi(C|y). 
\end{align*}
Choosing, say, a uniform prior for $N$ or a Jeffrey's prior, leads to different posterior inference for $C$. How then should we specify a prior for $N$?

Ideally, we would like to obtain a reparameterization $\theta'=[C^T,{N'}^T]^T$ such that $C$ and $N'$ are independent. Without loss of generality, first assume $N$ is a scalar.  Suppose we adopt flat priors on $C$ and $N$, and obtain $n$ samples from the corresponding posterior. Let $\tilde{C}$ be the $n\times p$ matrix of samples of $C$ and $\tilde{N}$ the $n-$vector of samples of $N$.  To obtain a transformed $\tilde{N'}=A \tilde{N}$ such that $\tilde{C}^T \tilde{N'}=0$, one could simply take the orthogonal complement to the projection of $\tilde{N}$ onto the column space of $\tilde{C}$. That is,
\begin{align}
\tilde{N'}&=\tilde{N}-\tilde{C}\hat{{\beta}}, \\
\text{where }~~ \hat{{\beta}}&=(\tilde{C}^T \tilde{C})^{-1}\tilde{C}^T \tilde{N}.
\end{align}
Note that $\tilde{N'}$ is equivalent to the vector of residuals from an ordinary least squares fit of the linear model 
\begin{align}
    E[N]=C^T {\beta}, \label{eqn:linmod}
\end{align} and it is easily verified that  $\tilde{C}^T \tilde{N'}=0$. This motivates the orthogonal parameterization:
\begin{align}
N'=N-{C}^T\hat{{\beta}},
\label{eq:leastsquare_transform}
\end{align}
which is the projection of $N$ onto the orthogonal complement of the column space of $\tilde{C}$. 

\subsubsection*{Non-Linear and Non-Additive Relationships}
\label{subsec:GAM}
While this yields a reparameterization $\theta^*$ such that $N'$ is approximately \textit{uncorrelated} each of the parameters in $C$, correlation represents only a specific linear form of dependence; this implies \textit{independence} between $N'$ and $C$ only if they are multivariate normally distributed\footnote{in this case there are no projection effects.}. In principle we could extend the above approach to ensure that $N'$ is uncorrelated with some non-linear function of $C$. Instead of fitting the standard linear model (\ref{eqn:linmod}), for example, we could regress $N$ on both $C_l$ and $C_l^2$ for each $l=1,\dots,p$. That is, we fit the model $E[N|C]=[C_1,C_1^2,\dots,C_p,C_p^2]{\beta}$ via least squares and obtain $N'=N-[C_1,C_1^2,\dots,C_p,C_p^2] \hat{\beta}$. This ensures $N'$ is uncorrelated with $C_l$ as well as $C^2_l$, $l=1,\dots,p$.

Rather than pre-specify the functional relationship between $N$ and $C_l$ as linear, quadratic, or some other parametric form, we instead estimate this functional form in a data-adaptive way, as in a Generalized Additive Model (GAM). First approximate a smooth function $f_l(C_l)$ via spline basis expansion: $f_l(C_l) \approx \sum_{k}^K b_{lk}(C_l) \beta_{lk}$ where $b_{lk}(\cdot)$ are \textit{known} basis functions and $\beta_k$ are unknown parameters. Then one can fit the following model
\begin{align}
E[N|C]=\sum_{l}^p f_l(C_l) \approx \sum_{l}^p \sum_{k}^K b_{lk}(C_l) \beta_{lk}=b^T \beta
\label{eq:GAM_transform}
\end{align}
where $b$ is the known vector of elements $b_{lk}(C_l)$ and $\beta$ is the corresponding vector of $p\times K$ unknown coefficients. This is again a linear model---albeit a much more flexible one than the ones above---and hence one can again fit this via least squares to obtain $\hat{\beta}=(\tilde{B}^T \tilde{B})^{-1}\tilde{B}^T\tilde{N}$ , yielding the orthogonal parameterization $N'=N-b^T\hat{\beta}$, which is approximately uncorrelated with each $f_l(C_l)$.

In practice ordinary least squares can lead to overfitting due to the extremely flexible basis representation. Instead, better stability can be achieved by penalizing overly ``wiggly'' functions. Specifically, one can encourage smoothness via a penalized least squares with quadratic penalty $\lambda \beta^T \Sigma \beta$ where $\Sigma$ is a known matrix corresponding to the the integrated second derivatives of the specified basis functions \citep{wood:2006}. This replaces the ordinary least squares estimator $\hat{\beta}$ with  $\hat{\beta}_{PLS}=(\tilde{B}^T \tilde{B}+\lambda \Sigma)^{-1}\tilde{B}^T \tilde{N}$. Here $\lambda$ is a penalty term that needs to be chosen; several methods such as generalized cross validation and restricted maximum likelihood are implemented in common software (e.g., {\tt pyGAM}\footnote{\url{https://github.com/dswah/pyGAM.git}}\cite{PyGAM} in python). 

This approach allows for arbitrary non-linear relationships between $N$ and $C_l$, but it assumes additivity among the $C_l$. One could in principle extend this argument further and fit non-linear and non-additive models for $N$ to allow for even more flexible dependence of $N$ on multiple cosmological parameters that may interact non-additively. This can be done by using appropriate tensor product basis expansions and similar penalized least squares estimators \cite{wood:2006}.

\subsubsection*{Multiple Nuisance Parameters}
Now suppose $N$ is a vector of length $p'$. A natural extension to the above approach would be to replace the univariate linear and additive models with multivariate regression models. However, this is not actually necessary in this case, as we need not make valid inferences with respect to $N$ at this phase; rather we only need to obtain an orthogonal reparameterization. Hence the above procedure can be straightforwardly repeated for each of $N_j$, $j=1,\dots,p'$. Similarly, we need not ensure the resulting parameters $N'_j$ are orthogonal to one another; as long as they are each orthogonal to $C$, then inference with respect to the cosmological parameters will be unaffected. 

Intuitively, some of of the variability in the original nuisance parameters $N$ is also captured by the cosmological parameters $C$. Because of this redundancy,  inferences about $C$ are informed not only by the prior for $C$ and the likelihood---but also the prior for $N$. The proposed approach instead reparameterizes the model in terms of $C$, the original target of inference, and $N'$, which only represents variation not already explained by $C$. This does not discard any information; it simply decouples the variation explained by cosmological parameters from that not explained by cosmological parameters. Inferences about $C$ now depend only on the likelihood and the prior for $C$---and is unaffected by choice of prior for $N'$.

\subsection{Outline of Proposed Method}
\label{proposed-method}
The proposed approach can be summarized in the following steps:
\begin{enumerate}
    \item \textit{Preliminary fit of cosmological model.}  Fit cosmological model with broad priors on $\{C,N\}$. Obtain posterior samples  $\{C_s,N_s\}$ for $s=1,\dots,S$.
    \item \textit{Orthogonalization model fit.} Using posterior samples $\{C_s,N_s\}$, fit a  model for $E[N|C]\approx B(C)\beta$ via penalized least squares to obtain estimate $\hat{\beta}$. This yields a transformation of parameters $\{C,N\}$ to $\{C,N'\}$, where $N'=[N-B(C)\hat{\beta}]$.
    \item \textit{Re-parameterized cosmological model fit.} Place meaningful physical priors on $C$ and some standard normal priors on the rescaled quantity $\tilde{N}'=N'/\alpha$. Then fit the re-parameterized cosmological model by transforming back the rotated $N'$ parameters through $N=[\alpha \tilde{N}'+B(C)\hat{\beta}] = [N'+B(C)\hat{\beta}]$ in the likelihood, with $\alpha$ a scaling factor estimated by sampling a constant likelihood with standard priors in the un-rotated $\{C,N\}$ basis, and projecting the parameter space onto $\{C,N'\}$. Obtain posterior samples  $\{C_s,N'_s\}$ for $s=1,\dots,S$.
\end{enumerate}
\subsection{Demonstration of Proposed Method in a Simple Example}

\begin{figure}
    \centering
    \includegraphics[width=0.49\linewidth]{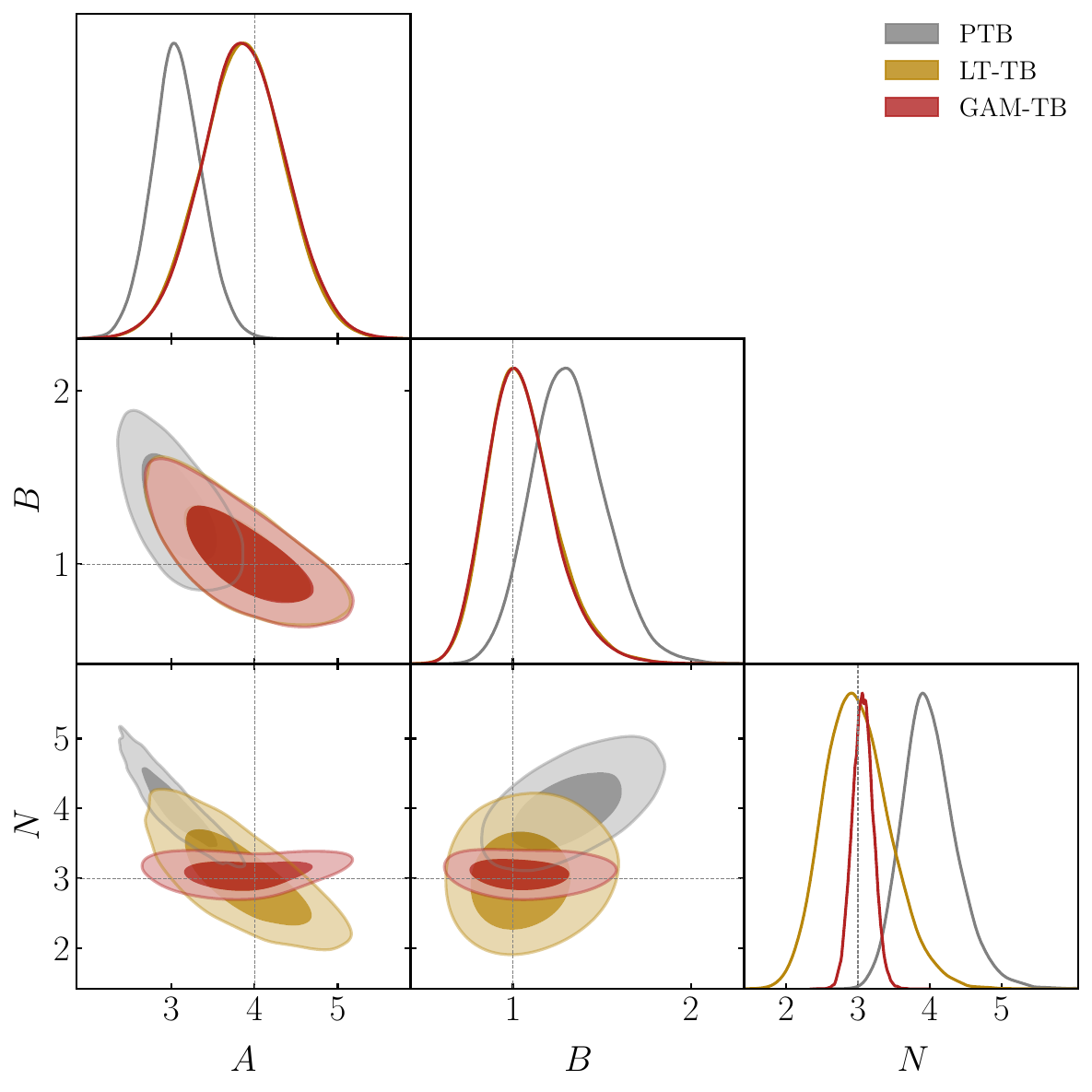}
    \includegraphics[width=0.49\linewidth]{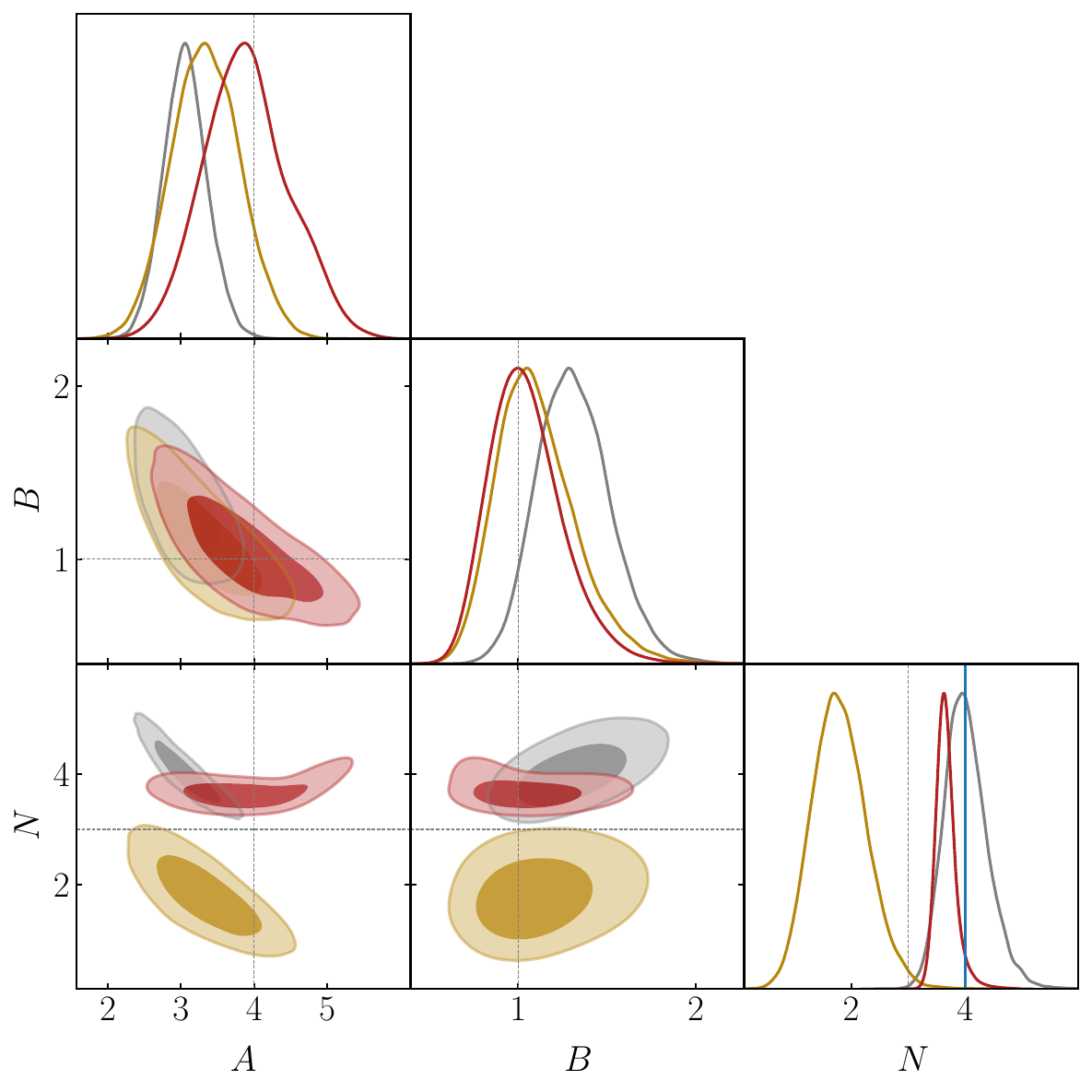}
    \caption{Posterior distribution for A, B and N from samples in the pre-transformed basis (PTB, grey), in the linear theory transformed basis (LT-TB, yellow) and non-linear transformed basis (GAM-TB, red). Left panel: we imposed full flat priors in the TB. Right panel: we imposed a skewed Gaussian prior on $N'$ centered at the blue vertical line.}
    \label{fig:toymodel}
\end{figure}

We demonstrate the potential of the proposed methodology through a simple toy model. We generated three data points ${y_1,y_2,y_3}$ according to the model: $y_1 = A$, $y_2=A\cdot B$ and $y_3=A\cdot N$; in this context, we treat A and B as our parameters of interest and N as the nuisance parameter, and model the log-likelihood according to a $\chi^2$ with covariance $C_{\rm ij}=\delta_{\rm i,j}\cdot\sigma$, where $\sigma=0.5$. 

We first fit a standard analysis to estimate the posterior for the original parameters $\{A,B,N\}$, adopting flat priors for target parameters $A$ and $B$. We also imposed a skewed normal prior on $N$ peaking $\sim 1\sigma$ away from the true (data-generating) value to induce a strong projection effect on $A$ and $B$. Resulting posteriors are displayed in grey in Fig.~\ref{fig:toymodel}. In the pairwise joint posterior plots of $\{A,N\}$ and $\{B,N\}$, we observe that the target parameters are highly correlated with the nuisance parameter $N$. Because of this, the prior on $N$ results in bias relative to the true values of $A$ and $B$ as seen in the marginal posterior plots.  

We then applied the proposed method (Steps 1--3) with two different orthogonalization models: a linear model (LM) fit via least squares, and a non-linear GAM, fit via penalized least squares. This results in two different re-parameterizations $\{C,N \}\rightarrow \{C,N'\}$, and we plot the resulting posteriors in yellow (LM) and red (GAM) in Fig.~\ref{fig:toymodel}. Results show the less flexible LM does not completely remove the dependence between $A$ and $N$, and the resulting posteriors of $A$ and $B$ still depend on the prior choice on $N$, as indicated by the right panel of the figure, where a skewed and off-centered Gaussian prior is imposed on $N'$. In contrast, the GAM transformation effectively decouples the parameters of interests from $N$, as demonstrated by the first and second lower 2D blocks of the triangular plot, for both the prior choices on $N'$ (left and right panels). Moreover, despite both the LM and the GAM orthogonalization models having been fit on samples that exhibit the projection effect, sampling in the transformed parameter spaces removed the bias induced on $A$ and $B$ by the prior on $N$. A noteworthy aspect of this test is the impact of the reparameterization on the posterior distribution width for N: it exhibits a smaller uncertainty and it is related the reduced volume of the joint posterior distribution once most of the correlation between A,B and N has been reduced--this can be seen as an effect of the A-N 2D posterior distribution rotation aligning to the axes, effectively yielding smaller uncertainties on the marginalized posterior of N, and a larger uncertainty on the marginalized posterior of A.

\subsection{Related Work}
The notion of parameter orthogonality was first formalized in a frequentist context by \cite{cox-reid-1987}, who defined it as having a diagonal information matrix \citep[see also][]{jeffreys-1961}. When the original parameters are not orthogonal to each other, one can find a transformation $\phi(\cdot,\cdot)$ of the original parameters $(\alpha, \beta)$, such that $\alpha'=\phi(\alpha, \beta)$ is orthogonal to $\beta$. However, finding $\phi$ requires solving differential equations \citep{cox-reid-1987}, which can be challenging with a large number of parameters. Simpler methods leading to approximate orthogonality have been proposed in the Bayesian context, e.g., the graphical approach in \cite{albert-1989} and the approximate transformation algorithm in \cite{tibshirani-wasserman-1994}. These transformations are helpful for reparameterizing the models to mitigate the influence of nuisance parameters on the estimation of the parameters of interest, i.e., the inference about $\beta$ is not significantly influenced by knowledge about $\alpha'$. 

It was noted by \cite{albert-1989} that if the joint posterior density of $\alpha$ and $\beta$ is concentrated in an ellipse, then their dependence can be well captured by the slope $\nu=\mathrm{cov}(\alpha,\beta)/\mathrm{var}(\beta)$. Consequently, a straightforward linear transformation to orthogonalize the parameters is $\alpha'=\alpha-\nu\beta$. This transformation is attractive in practice due to the ease of computation and implementation. Similarly, \cite{papaspiliopoulos-etal-2007} recommended a family of transformations $\alpha' = \alpha - \psi(\beta)$, for which even the simple linear transformation $\psi(x)=\nu x$ can be flexible enough for posterior orthogonalization of the parameters $(\alpha,\beta)$. In particular, they showed that this reparameterization leads to faster  MCMC convergence when $\alpha$ and $\beta$ are strongly dependent \textit{a priori} in the presence of insufficiently informative data, also known as the weak data case. The benefits of this parameterization method were also demonstrated in \cite{christensen-etal-2006}. We consider a more general non-linear transformation $\psi$ such that $\alpha=\psi(\beta)$, and approximate it via a GAM fit to the posterior samples from a preliminary model fit.

\section{Effective Field Theory}
\label{sec:EFT}
The EFTofLSS framework provides a comprehensive approach to modelling the redshift-space galaxy power spectrum, addressing the impact of small-scale physics on large-scale clustering. By systematically introducing counterterms, the EFTofLSS goes beyond standard perturbation theory to account for small-scale effects, including galaxy formation processes. A summary of the model is provided here, and we refer readers to ~\cite{Perko2016BiasedStructure,dAmico2020TheStructure} for further details.

The one-loop EFT model for the redshift-space galaxy power spectrum is expressed as:

\begin{align}\label{eqn:gpk}
 P_{g}(k, \mu) & = Z_1(\mu)^2 P_{11}(k) + 2 \int \frac{d^3q}{(2\pi)^3}\; Z_2(\mathbf{q},\mathbf{k}-\mathbf{q},\mu)^2 P_{11}(|\mathbf{k}-\mathbf{q}|)P_{11}(q)\nonumber  \\
& + 6 Z_1(\mu) P_{11}(k) \int\, \frac{d^3 q}{(2\pi)^3}\; Z_3(\mathbf{q},-\mathbf{q},\mathbf{k},\mu) P_{11}(q)\nonumber \\
& + 2 Z_1(\mu) P_{11}(k)\left( c_\text{ct}\frac{k^2}{{ k^2_\textsc{m}}} + c_{r,1}\mu^2 \frac{k^2}{k^2_\textsc{r}} + c_{r,2}\mu^4 \frac{k^2}{k^2_\textsc{r}} \right)\nonumber \\
& + \frac{1}{\bar{n}_g}\left( c_{\epsilon,0}+c_{\epsilon,1}\frac{k^2}{k_\textsc{m}^2} + c_{\epsilon,2} f\mu^2 \frac{k^2}{k_\textsc{m}^2} \right) \,
\end{align}

We adopt the notation of \cite{DAmico2021LimitsCode}, combining linear terms, 1-loop Standard Perturbation Theory (SPT) terms, counterterms, and stochastic contributions. Here, $\mu$ refers to the cosine of the angle between the line of sight and the wavenumber vector $\mathbf{k}$. The linear matter power spectrum is represented by $P_{11}(k)$, and $f$ denotes the growth factor. The scales $k_\textsc{m}^{-1}$ and $k_\textsc{r}^{-1}$ define the spatial extension of collapsed objects and counterterms related to velocity products at a point, respectively \cite{DAmico2024TamingData,Ivanov2022CosmologicalDistortions}. The mean galaxy number density is denoted as $\bar{n}_g$. The $Z_n$ are the $n$-th order galaxy density kernels, defined as:

\begin{align}\label{eqn:zkernels}
    Z_1(\mathbf{q}_1) & = K_1(\mathbf{q}_1) +f\mu_1^2 G_1(\mathbf{q}_1) = b_1 + f\mu_1^2\,, \nonumber\\ 
    Z_2(\mathbf{q}_1,\mathbf{q}_2,\mu) & = K_2(\mathbf{q}_1,\mathbf{q}_2) +f\mu_{12}^2 G_2(\mathbf{q}_1,\mathbf{q}_2)+ \, \frac{1}{2}f \mu q \left( \frac{\mu_2}{q_2}G_1(\mathbf{q}_2) Z_1(\mathbf{q}_1) + \text{perm.} \right)\,, \nonumber\\ 
    Z_3(\mathbf{q}_1,\mathbf{q}_2,\mathbf{q}_3,\mu) & = K_3(\mathbf{q}_1,\mathbf{q}_2,\mathbf{q}_3) + f\mu_{123}^2 G_3(\mathbf{q}_1,\mathbf{q}_2,\mathbf{q}_3) \nonumber \\ 
    &+ \frac{1}{3}f\mu q \left(\frac{\mu_3}{q_3} G_1(\mathbf{q}_3) Z_2(\mathbf{q}_1,\mathbf{q}_2,\mu_{123}) +\frac{\mu_{23}}{q_{23}}G_2(\mathbf{q}_2,\mathbf{q}_3)Z_1(\mathbf{q}_1)+ \text{cyc.}\right)\,,
\end{align}

Here, $K_n$ refers to the $n$-th order galaxy density kernels, expressed as:

\begin{align}\label{eqn:kkernels}
    K_1 & = b_1\,, \nonumber\\
    K_2(\mathbf{q}_1,\mathbf{q}_2) & = b_1\frac{\mathbf{q}_1\cdot \mathbf{q}_2}{q_1^2}+ b_2\left( F_2(\mathbf{q}_1,\mathbf{q}_2)- \frac{\mathbf{q}_1\cdot \mathbf{q}_2}{q_1^2} \right) + b_4 + \text{perm.} \,, \nonumber\\
    K_3(k, q) & = \frac{b_1}{504 k^3 q^3}\left( -38 k^5q + 48 k^3 q^3 - 18 kq^5 + 9 (k^2-q^2)^3\log \left[\frac{k-q}{k+q}\right] \right) \nonumber\\
    &+ \frac{b_3}{756 k^3 q^5} \left( 2kq(k^2+q^2)(3k^4-14k^2q^2+3q^4)+3(k^2-q^2)^4 \log \left[\frac{k-q}{k+q}\right] \nonumber \right) \,.
\end{align}

For brevity, the explicit forms of the second-order density kernel $F_2$ and the velocity kernels $G_n$ from standard perturbation theory are not included here. To summarize, the model uses four galaxy bias parameters: $b_1, b_2, b_3$, and $b_4$, three counterterms: $c_{\rm{ct}}, c_{r,1}$, and $c_{r,2}$, and three stochastic parameters: $c_{\epsilon,0}, c_{\epsilon,1}, c_{\epsilon,2}$. Altogether, these yield a total of 10 nuisance parameters: \begin{align} { b_1, b_2, b_3, b_4, c_{\rm{ct}}, c_{r,1}, c_{r,2}, c_{\epsilon,0}, c_{\epsilon,1}, c_{\epsilon,2} },. \end{align} The power spectrum is IR-resummed at one loop order to address the effects of long-wavelength modes that could otherwise lead to divergences in perturbative calculations~\cite{Senatore2014RedshiftStructures, Senatore2015TheStructures, Lewandowski2020AnPeak}.

\section{Simulation pipeline}
\label{sec:dataset}

In this section, we describe the simulation pipeline designed to evaluate the impact of our proposed coordinate transformation in the EFTofLSS framework. As introduced in Sec.~\ref{sec:Intro}, the low signal-to-noise ratio does not allow us to decouple the EFT parameters from the cosmological ones, effectively yielding bias for the latter due to the impact of the prior choice on the EFT parameters -- also referred to as \textit{projection effects}. 

We study the effectiveness of the proposed reparametrization in a statistical ensemble. We therefore generated 100 cosmological datasets in the following way: for each realization, we generated galaxy monopole and quadrupole power spectra at redshift $z=0.5$, up to a maximum wave number $k_{\rm max} = 0.25 \, h/\text{Mpc}$. The fiducial cosmological model employed in this study is shown in Table~\ref{tab:fiducial_cosmology}. 
Noise realizations were drawn from a covariance matrix obtained from the \texttt{CovaPT} code~\cite{Wadekar:2019rdu}; we used the same covariance employed in~\cite{Zhang:2024}, which corresponds to a galaxy density of $5 \times 10^{-4} \mathrm{Mpc}^{-3}$ and a volume of $8\,(\mathrm{Gpc}/h)^3$

\begin{table}[htbp!]
    \footnotesize
    \centering
    {\renewcommand{\arraystretch}{1.2}
    \begin{tabular}{l c r}
    \hline
    Parameter name  & Fiducial value & Description \\
    \hline
    \hline
    $\ln 10^{10}A_{\rm s}$                                  & 3.044   & Amplitude of scalar perturbations ($k_{0}=0.05\rm{Mpc}^{-1}$)\\
    $n_{\rm s}$                                             & 0.965   & Index of the scalar perturbation spectrum \\
    $h \equiv \rm{H}_0/100 $                           & 0.675   & Reduced Hubble parameter \\
    $\rm m_\nu$                                             & 0.06    & Neutrino mass [eV] \\
    $\omega_{\rm b}\equiv \Omega_{\rm b}h^2$           & 0.02235 & Baryon physical density parameter times $\rm h^2$ \\
    $\omega_{\rm c}\equiv \Omega_{\rm c}h^2$           & 0.120   & Cold dark matter physical density parameter times $\rm h^2$ \\
    $w_0$                                                   & -1      & Dark energy equation of state parameter \\
    \hline
    \end{tabular}}
    \caption{Fiducial cosmology parameters used for generating power spectra realizations.}
    \label{tab:fiducial_cosmology}
\end{table}

\begin{table}[htbp!]
    \footnotesize
    \centering
    {\renewcommand{\arraystretch}{1.2}
    \begin{tabular}{l c r}
    \hline
    Parameter name  & Fiducial value & Standard prior\\
    \hline
    \hline
    $b_{1}$           & $2.0$ & $[0,4]$             \\
    $b_{2}$           & $1.0$ & $[-4,4]$            \\
    $b_{3}$           & $0$   & $\mathcal{N}(0,10)$ \\
    $b_{4}$           & $0$   & $\mathcal{N}(0,2)$  \\
    $c_{\rm ct}$      & $0$   & $\mathcal{N}(0,4)$  \\
    $c_{\rm r, 1}$    & $0$   & $\mathcal{N}(0,2)$  \\
    $c_{\rm r, 2}$    & $0$   & fixed               \\
    $c_{\epsilon, 0}$ & $0$   & $\mathcal{N}(0,2)$  \\
    $c_{\epsilon, 1}$ & $0$   & fixed               \\
    $c_{\epsilon, 2}$ & $0$   & $\mathcal{N}(0,4)$  \\
    \hline
    \end{tabular}}
    \caption{Fiducial EFT parameters used for generating power spectra realizations with standard prior used in the PTB runs.}
    \label{tab:fiducial_EFT}
\end{table}

For each of the 100 datasets, we applied the proposed methodology as described in Section \ref{sec:Methods}. In the preliminary cosmological fit (Step 1), we adopted priors outlined in Table~\ref{tab:fiducial_EFT} (we choose $k_M = 0.7 h/\textrm{Mpc}$, $k_R = 0.35 h/\textrm{Mpc}$) ; we imposed flat priors on all the cosmological parameters and include information from BBN in the form of a Gaussian prior of $\Omega_bh^2=0.02235\pm0.00037$ under the assumption of $N_{\rm{eff}}=3.046$. This prior is derived using the empirically-estimated cross-section of the deuterium described in \cite{Adelberger:2011}, with abundance $\mathrm{D}/\mathrm{H} = (2.527\pm0.030)\times 10^{-5}$ from the high-resolution spectroscopic measurements of seven quasar absorption systems \cite{Cooke:2018}. 

In Step 2 we considered two different parameter transformations based on the following orthogonalization models: 
\begin{enumerate}
    \item[(a)] \textbf{Linear Model (LM)}: Here we fit a main effects linear model $E[N_j|\mathbf{C}]=\mathbf{C}\boldsymbol{\beta}_j$ via least squares. This minimizes the covariance between $\mathbf{C}$ and $\mathbf{N'}$ as described in Sec.~\ref{subsec:least-square}. 
    \item[(b)] \textbf{GAM}: To account for potential non-linear dependencies, we instead fit the following non-linear GAM model via penalized least squares: $$E[N_j|\mathbf{C}]=f_{1j}(\ln 10^{10}A_\mathrm{s}) + f_{2j}(n_\mathrm{s}) + f_{3j}(\omega_\mathrm{b}) + f_{4j}(w_0) + f_{5j}(\mathrm{h}) + f_{6j}(\omega_\mathrm{c}).$$  where the $f_{lj}$ indicate smooth functions approximated by penalized b-splines with $20$ basis functions of order $3$. 
\end{enumerate}

As a consistency check, we tested that the 100 GAMs computed for the Step 2 correctly map the true cosmological input parameters onto the input EFT parameters, by plugging the input cosmological parameter values (Tab.~\ref{tab:fiducial_cosmology} and Tab.~\ref{tab:fiducial_EFT}) into Eq.~\ref{eq:GAM_transform}; details of this test are reported in Appendix~\ref{app:GAM_predcheck}.

In the re-parameterized cosmological model fit (Step 3), we adopted standard normal priors on $N'=\left[b_1',b'_2,..\right]$ ($\mathcal{N}(0,1)$, we will refer to it as "\textit{standard normal priors}" onwards in the text) and estimate, for each of the 100 datasets, the scaling factor $\alpha$ for the rotated nuisance parameters in the likelihood. 

Although the scaling factor $\alpha$ is estimated to match the amplitude of PTB EFT parameters with standard priors in the PTB, we investigated how Standard Normal priors in the transformed basis (TB) map onto the pre-transformed basis (PTB) space. We therefore checked that such prior choices did not (1) cut out significant regions of the original  EFT parameter space or (2) result in prior-dominance for the PTB parameters. This has been verified by comparing the width of the induced priors with the width of the EFT standard priors in the PTB. Details of this validation are in Appendix~\ref{app:prior}.

We use \texttt{Effort.jl}~\cite{Bonici:2025ltp} to compute galaxy clustering multipoles as functions of cosmological and nuisance parameters. \texttt{Effort.jl} is a novel emulator for the EFTofLSS, based on \texttt{PyBird}~\cite{DAmico2021LimitsCode}, and shares its computational backend with \texttt{Capse.jl}~\cite{Bonici:2023xjk}. 
Given the large number of chains we are running, and our goal of achieving robust convergence even without analytical marginalization, we employ \texttt{Effort.jl}. 
The MCMC sampling procedure in Steps 1 \& 3 was carried out with the publicly available tool {\tt pocomc}\footnote{\url{github.com/minaskar/pocomc}}\cite{karamanis2022accelerating,karamanis2022pocomc}. 

\subsection{Iterative pipeline}

Our GAM fitting procedure is performed sequentially for each parameter. As such it is not guaranteed that this will reach the optimal solution across all parameters. 
A natural improvement of the methodology would be to iterate the procedure to remove residual dependence between cosmological and nuisance parameters at each step. Iterations are performed in the following way:
\begin{enumerate}
    \item for iteration $i$-th, we fit a GAM to the transformed samples from iteration $i-1$, resulting in a new transformation $B^{(i)}(C)\beta^{(i)}\approx E[N^{(i-1)}|C]$;
    \item we generate new samples in the rotated basis $\{C_{s},N^{(i)}_{s}\}$, for $s=1,...,S$ by exploiting the linearity of the transformation so that $N = \alpha^{(i)}N^{(i)} + \sum_{j=0}^{i}B^{(j)}(C)\beta^{(j)}$.
\end{enumerate}
An important aspect of this approach is that the GAM transformation becomes less constraining through iterations, and the resulting nuisance parameter volume covered by the transformation increases with further iterations. Heuristically, this can be seen as the parameters to be less and less correlated, and therefore the GAM transformation loses its predictive ability to map the cosmological parameters onto the EFT parameters space. This requires the rotated parameters scaling factors $\alpha$ in the likelihood to be adjusted at each step, allowing for a larger amplitude which in turns requires longer convergence time of the MCMC sampling to achieve convergence. For each iteration we perform the same procedure adopted for a single iteration: we computed, after step (1) and before step (2), the expected TB volume space by sampling a constant likelihood with standard priors in the PTB, and then applying the $i$-th GAM transformation; this allows estimation of scaling factors $\alpha$ that approximately match the standard prior amplitude in the PTB. 

\section{Results}
\label{sec:Results}
In this section we show our results after applying the presented methodology at first with a single iteration, and then after multiple iterations.
\subsubsection*{Single iteration results}

To assess the effectiveness of the proposed re-parameterization at reducing projection effects, we analyzed the posterior distributions of cosmological and EFT parameters across 100 realizations, comparing the results of the proposed approach to those of the preliminary  cosmological fits (Step 1 only), which represent a na\"ive analysis subject to any projection effects. We first report posteriors after marginalizing over sampling variability by stacking the 100 MCMC chains. We also report frequentist properties: for each parameter we plot sampling distributions of the maximum a posteriori (MAP) estimates across the 100 datasets, and we report also relative bias, in units of the standard deviation, and 95\% credible interval coverage.

\begin{figure}
    \centering
    \includegraphics[width=0.95\linewidth]{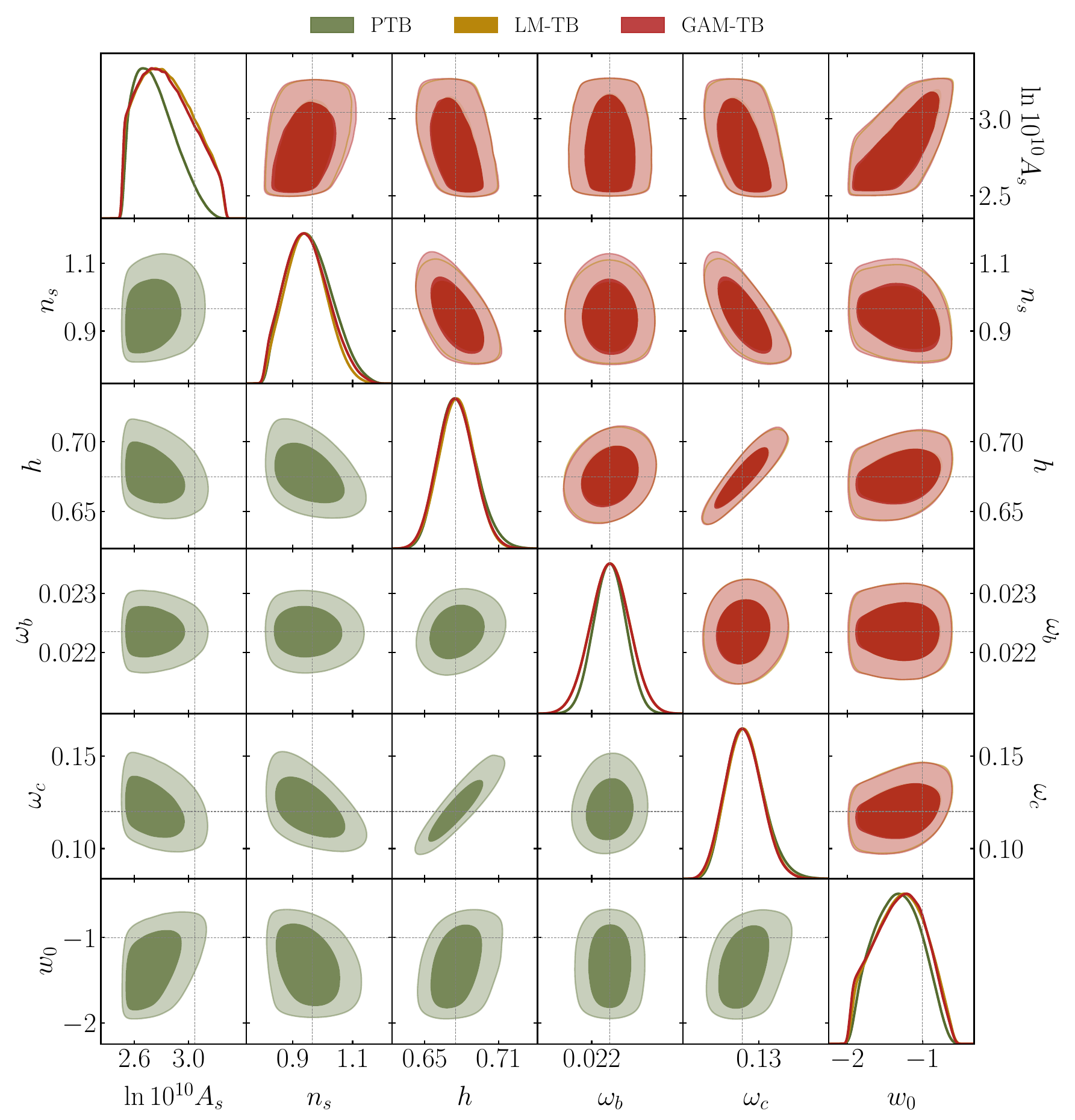}
    \caption{Posterior distributions of the $w_0$CDM parameters comparison between the PTB (green), the GAM-TB with regularization priors on the rotated parameters (red), the LM-TB with flat priors on the rotated parameters (yellow).}
    \label{fig:LCDM}
\end{figure}

\begin{sidewaysfigure}
    \centering
    \includegraphics[height=12cm]{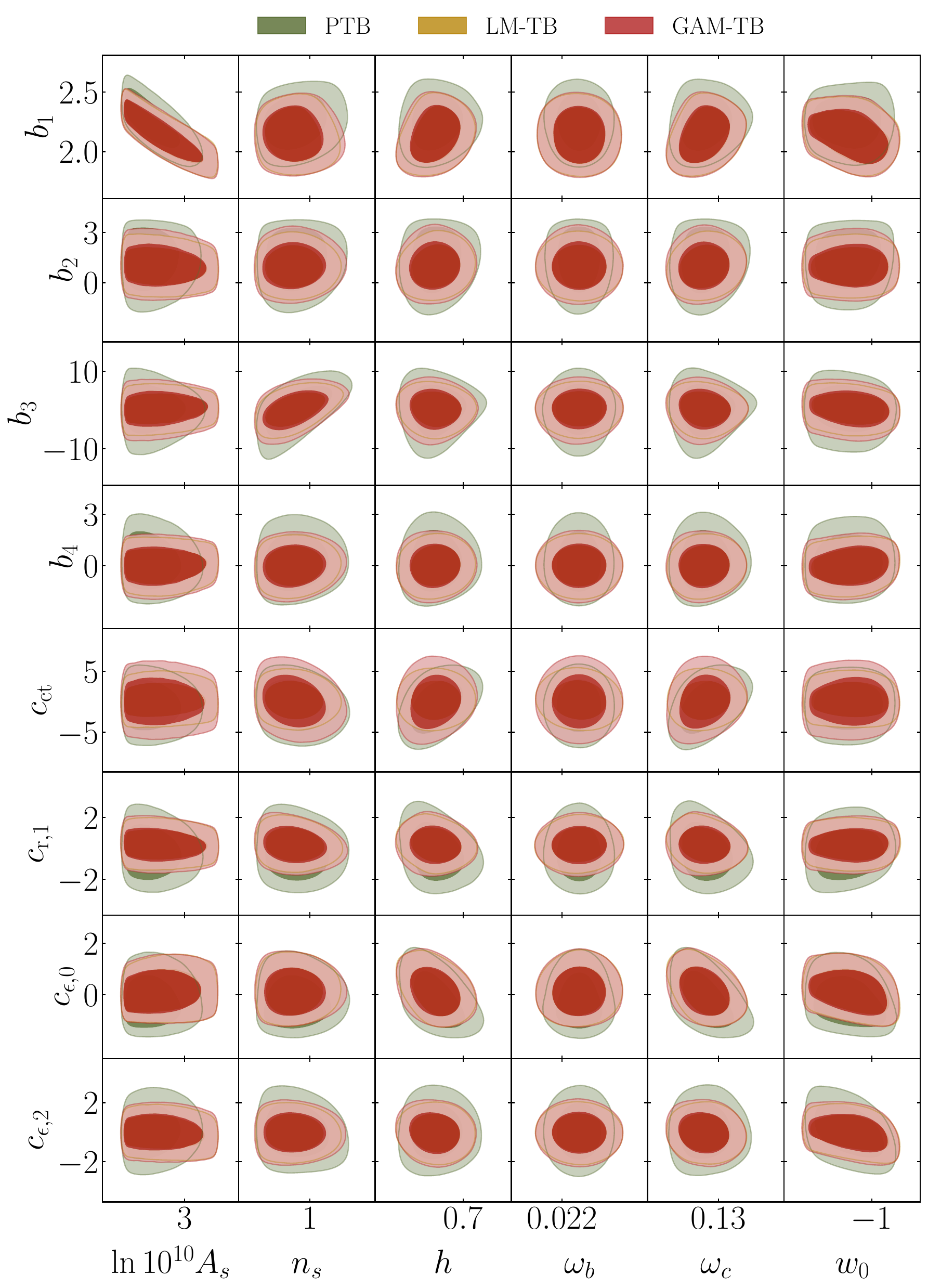} \qquad\includegraphics[height=11.5cm]{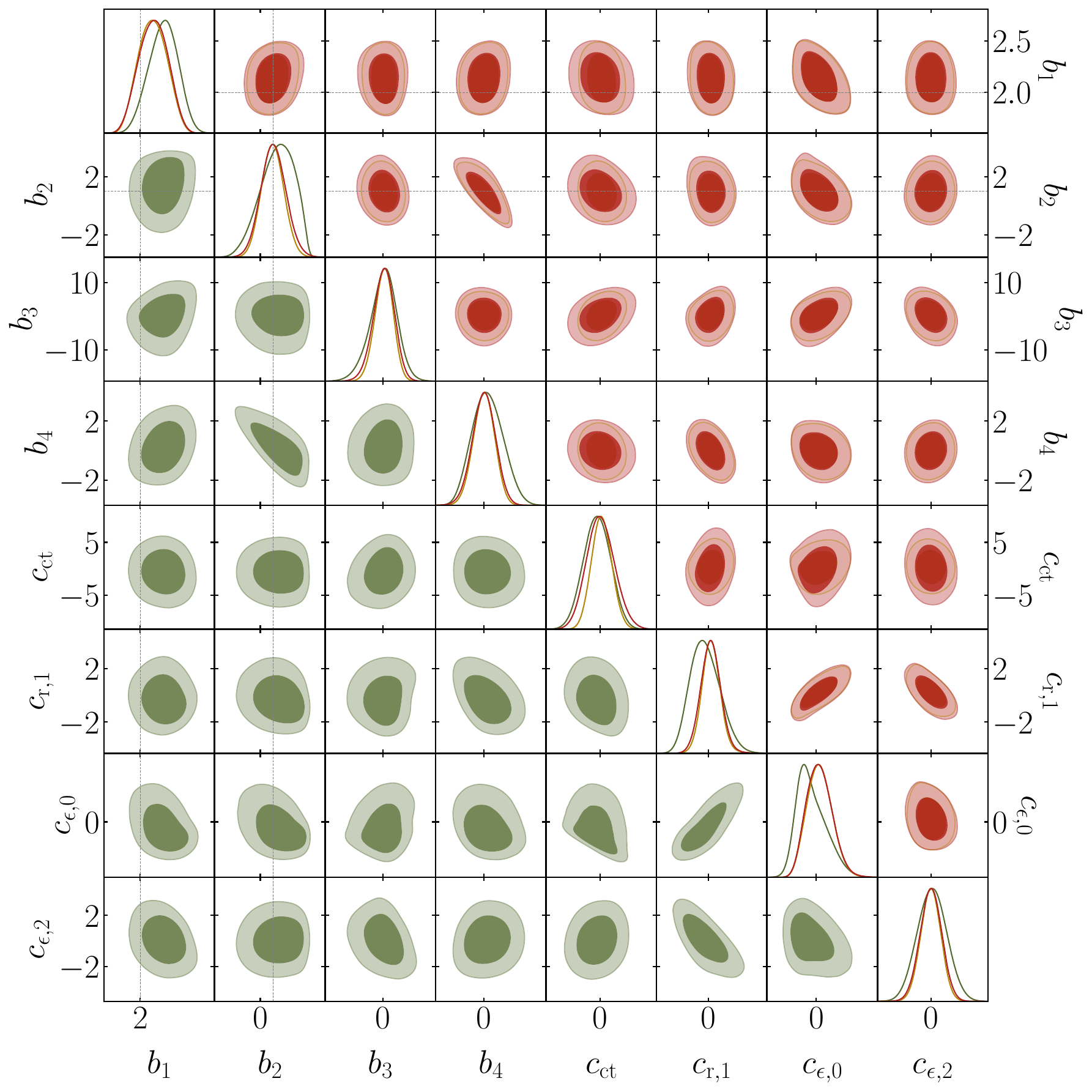}
    \caption{Analogous of Fig.~\ref{fig:LCDM} but showing the EFT parameters in the standard basis (right) and the 2D plots of cosmological and EFT parameters (left). For the MCMC runs in the rotated basis, we performed the inverse transformation to derive the posterior in the original EFT basis.}
    \label{fig:fullpars}
\end{sidewaysfigure}

\begin{figure}
    \centering
    \includegraphics[width=0.9\linewidth]{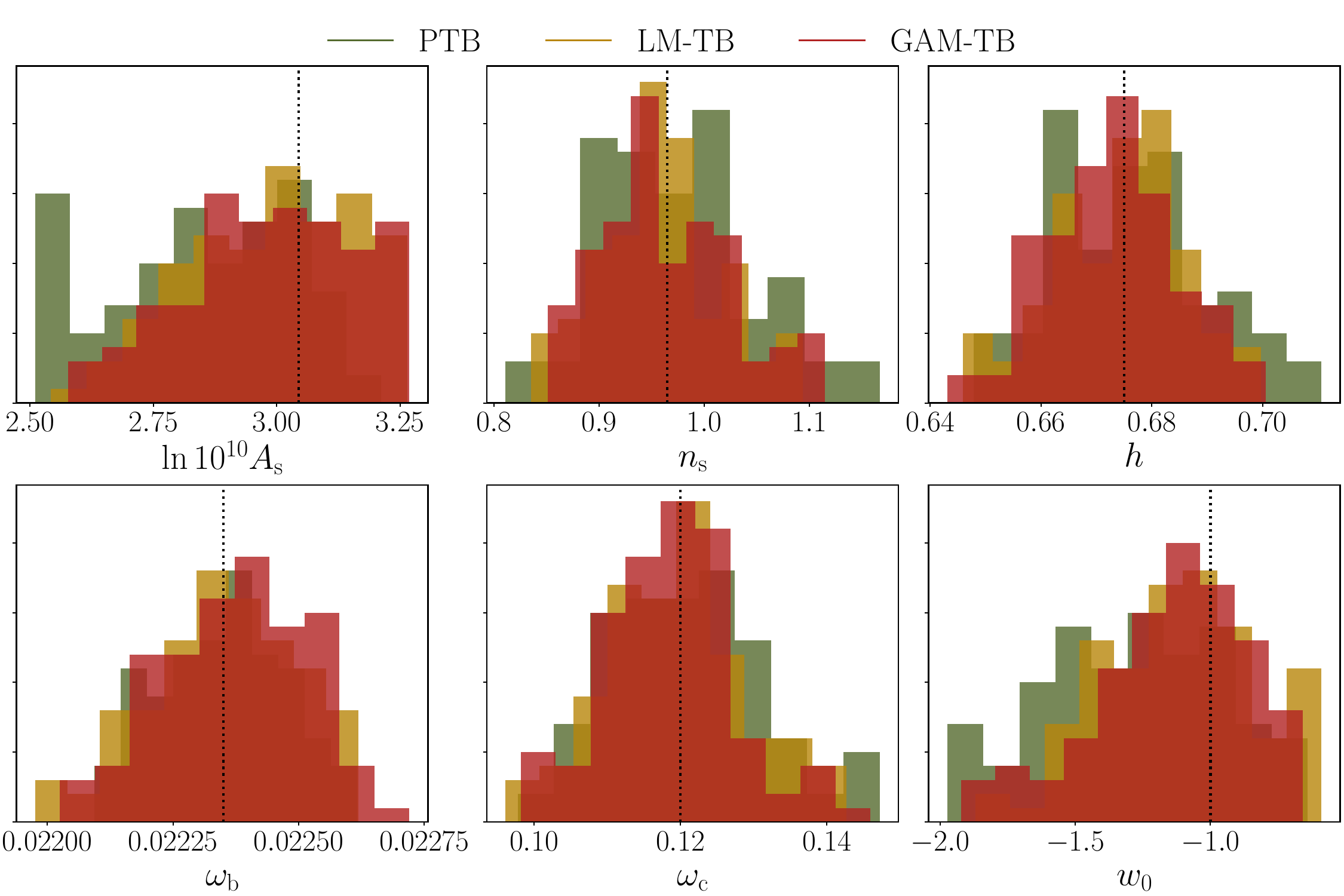} \\
    \caption{Frequentist properties of the 100 realization. The histograms show the distribution of the MAP values for the $w0$CDM parameters among the 100 realization considered in this work comparing the PTB (green), the LM-TB (yellow), and the GAM-TB (red).}
    \label{fig:LCDM_freqstats}
\end{figure}

\begin{figure}
    \centering
    \includegraphics[width=0.49\linewidth]{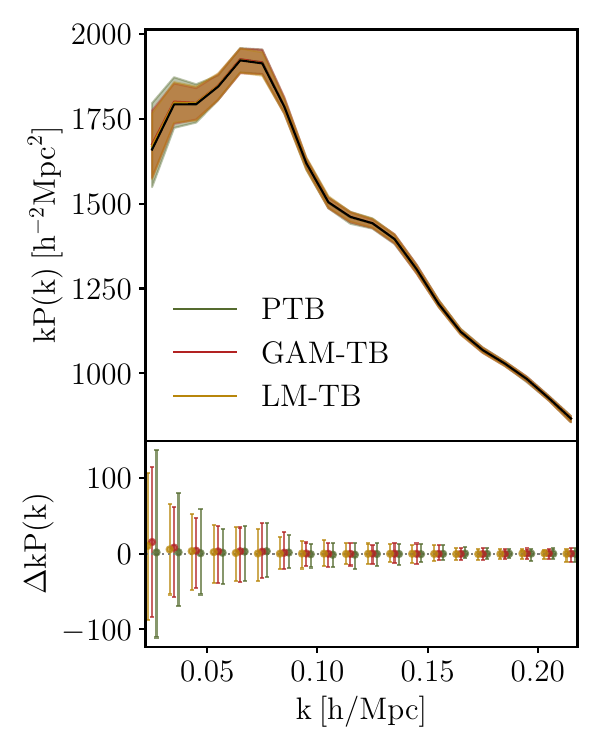}
    \includegraphics[width=0.49\linewidth]{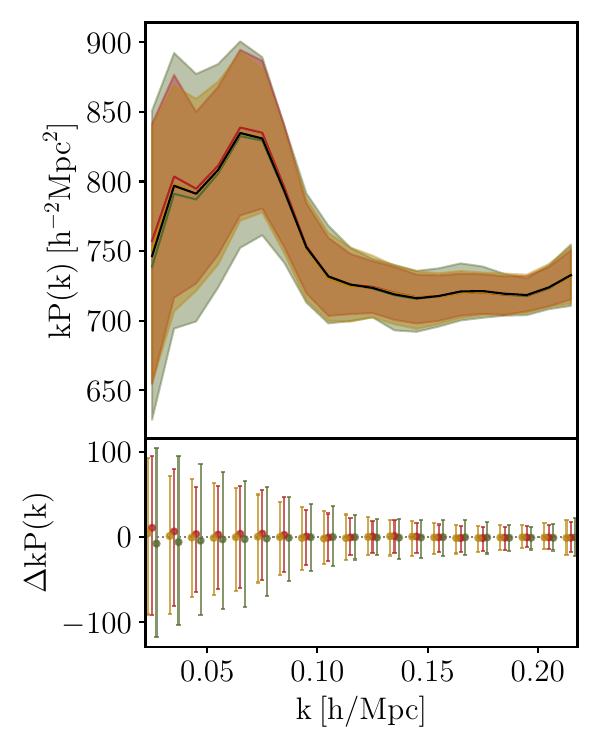}
    \caption{Best-fit monopole (left) and quadrupole (right) power spectra, reconstructed from the MCMC chains in the PTB (green), LM-TB (yellow), GAM-TB (red). The input power spectra are plotted in solid black lines. Coloured regions (are error bars) report 95\% CI.}
    \label{fig:bestfit_ps}
\end{figure}

We show in Fig.~\ref{fig:LCDM} the posteriors for the target cosmological parameters, and in Fig.~\ref{fig:fullpars} for the full set of $w0$CDM parameters and EFT parameters in the un-rotated basis, obtained by stacking the 100 MCMC chains. Our findings indicate that a single application of the proposed reparametrization  somewhat mitigates the projection effects in the marginal distributions of cosmological parameters, especially on the scalar amplitude $\ln 10^{10}A_{\rm s}$ and $w_0$. This improvement was consistently observed across realizations, as displayed in Fig.~\ref{fig:LCDM_freqstats}, illustrating the distribution of the maximum a posteriori (MAP) estimates for the cosmological parameters across the 100 simulated datasets. We similarly find that the less flexible linear-model based transformation performs comparably well at mitigating the projection effects (Tab.~\ref{tab:freq_res}). This suggests that the $C$--$N$ parameter coupling may be dominated by a linear relationship in the likelihood here, but this is not necessarily true in other models.

\begin{table}[htbp!]
    \footnotesize
    \centering
    {\renewcommand{\arraystretch}{1.2}
    \begin{tabular}{l|lcr|lcr}
    \hline
    & \multicolumn{3}{|c|}{$\sigma$} & \multicolumn{3}{c}{$\Delta\left[ \sigma\right]$} \\
    Parameter  & PTB & LM-TM & GAM-TB & PTB & LM-TM & GAM-TB \\
    \hline
    \hline
    $\ln 10^{10}A_{\rm s}$  & 0.19 & 0.17 & 0.17 & -1.1 & -0.3 & -0.3 \\
    $n_{\rm s}$             & 0.07 & 0.06 & 0.06 & 0.12 & 0.12 & 0.04 \\
    $h  $                   & 0.014 & 0.012 & 0.013 & -0.1 & -0.01 & -0.17 \\
    $\omega_{\rm b}$        & 0.00012 & 0.00014 & 0.00011 & -0.002 & 0.2 & 0.004 \\
    $\omega_{\rm c}$        & 0.01 & 0.01 & 0.01 & 0.1 & -0.05 & -0.1 \\
    $w_0$                   & 0.3 & 0.3 & 0.3 & -1.0 & -0.4 & -0.5 \\
    \hline
    \end{tabular}}
    \caption{Frequentist properties of PTB, LM-TB and GAM-TB estimators standard deviation and relative bias (in units of $\sigma$) of MAP estimates for the cosmological parameters across 100 datasets.}
    \label{tab:freq_res}
\end{table}

In addition, we evaluated the power spectra at the best-fit parameter values. As shown in Fig.~\ref{fig:bestfit_ps}, both the analyses in the PTB and the TB yielded power spectra that were consistent with the input spectra, even though both approaches reached similar minima in the joint log-posterior, with $\Delta\chi^2$ with respect to the PTB as small as $0.04$ for the LM-TB and $0.33$ for the GAM-TB. These results suggest that both the linear-model and the GAM-based transformations effectively decouple the cosmological parameters from the nuisance parameters, with a minimal impact on the accuracy of the power spectrum predictions.

\begin{figure}
    \centering
    \includegraphics[width=0.9\linewidth]{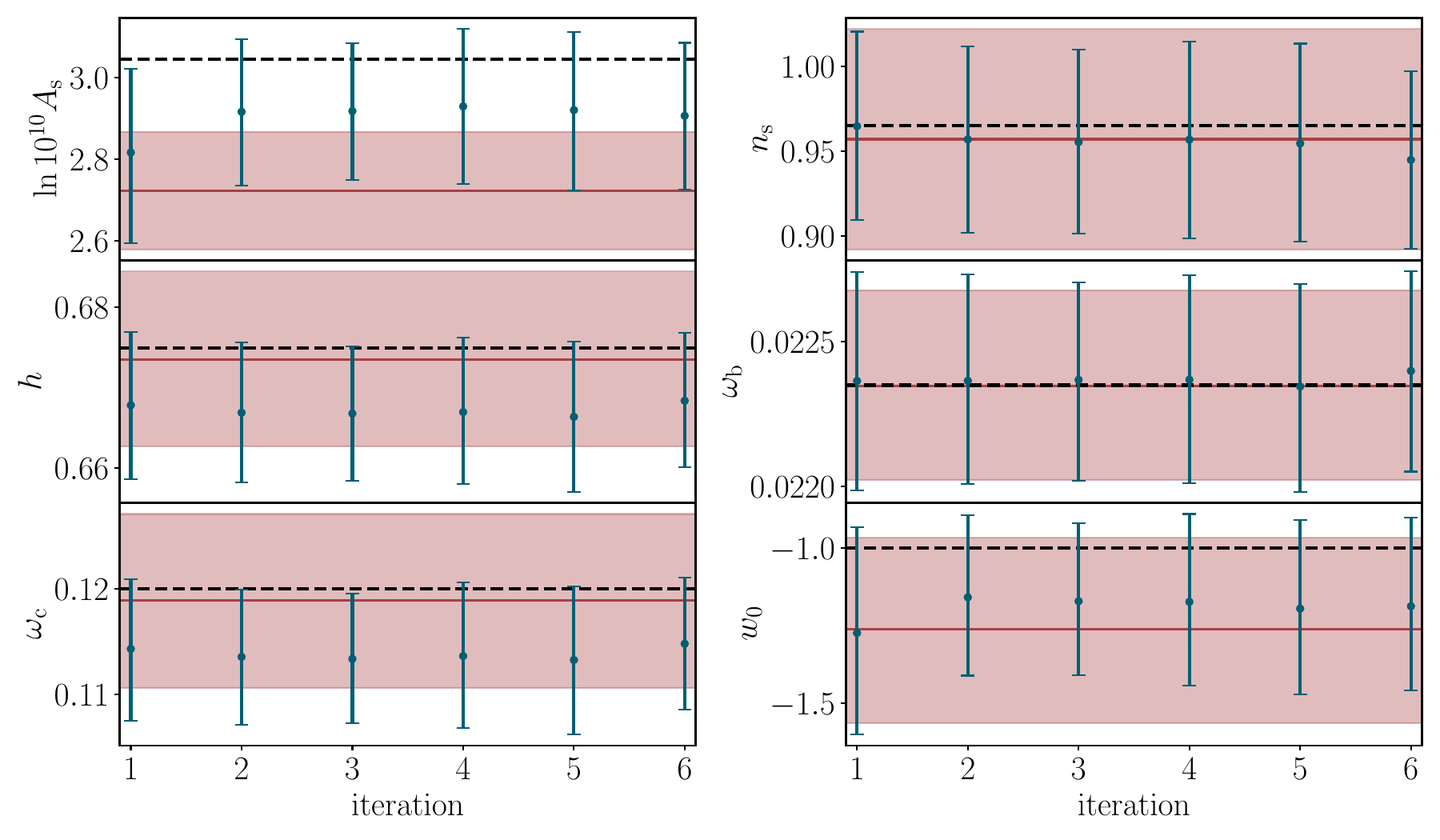} \\
    \caption{Cosmological parameters estimates (mean and 68\% CI) for six iterations of the GAM transformation on a single dataset. The red region corresponds to the 68\%CI from the posteriors in the PTB run. Black dashed lines report the true input value in our simulations.}
    \label{fig:LCDM_iter}
\end{figure}

\begin{figure}
    \centering
    \includegraphics[width=0.9\linewidth]{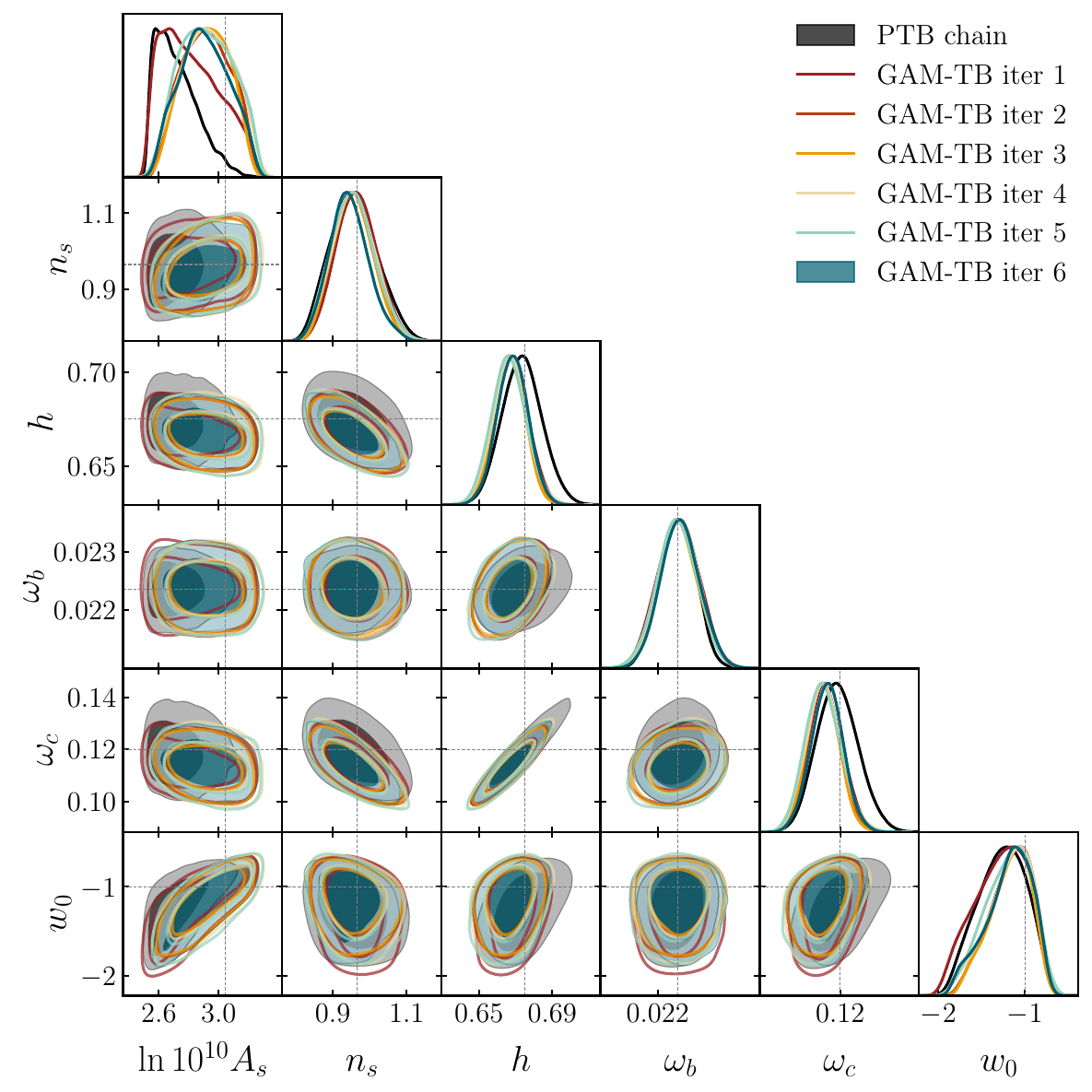} \\
    \caption{Triangle plot for six iterations of the GAM transformation on a single dataset corresponding to Fig.~\ref{fig:LCDM_iter}. The figure reports the PTB posterior (black), and iteration 1 (red), 2 (yellow) and 3 (teal) of our GAM NL iterative methodology.}
    \label{fig:LCDM_iter_triangle}
\end{figure}

\begin{sidewaysfigure}
    \centering
    \includegraphics[height=12cm]{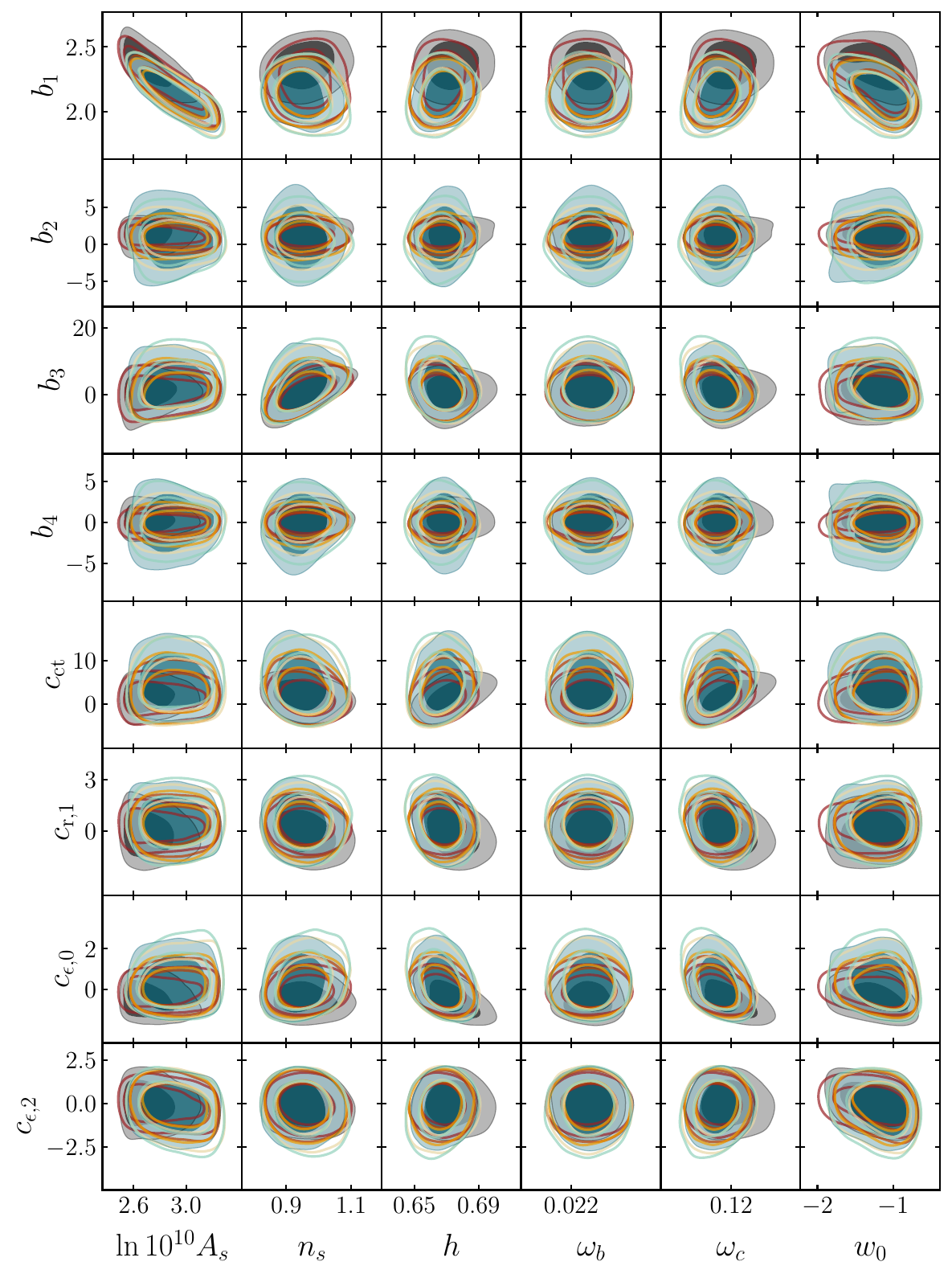} \qquad\includegraphics[height=12cm]{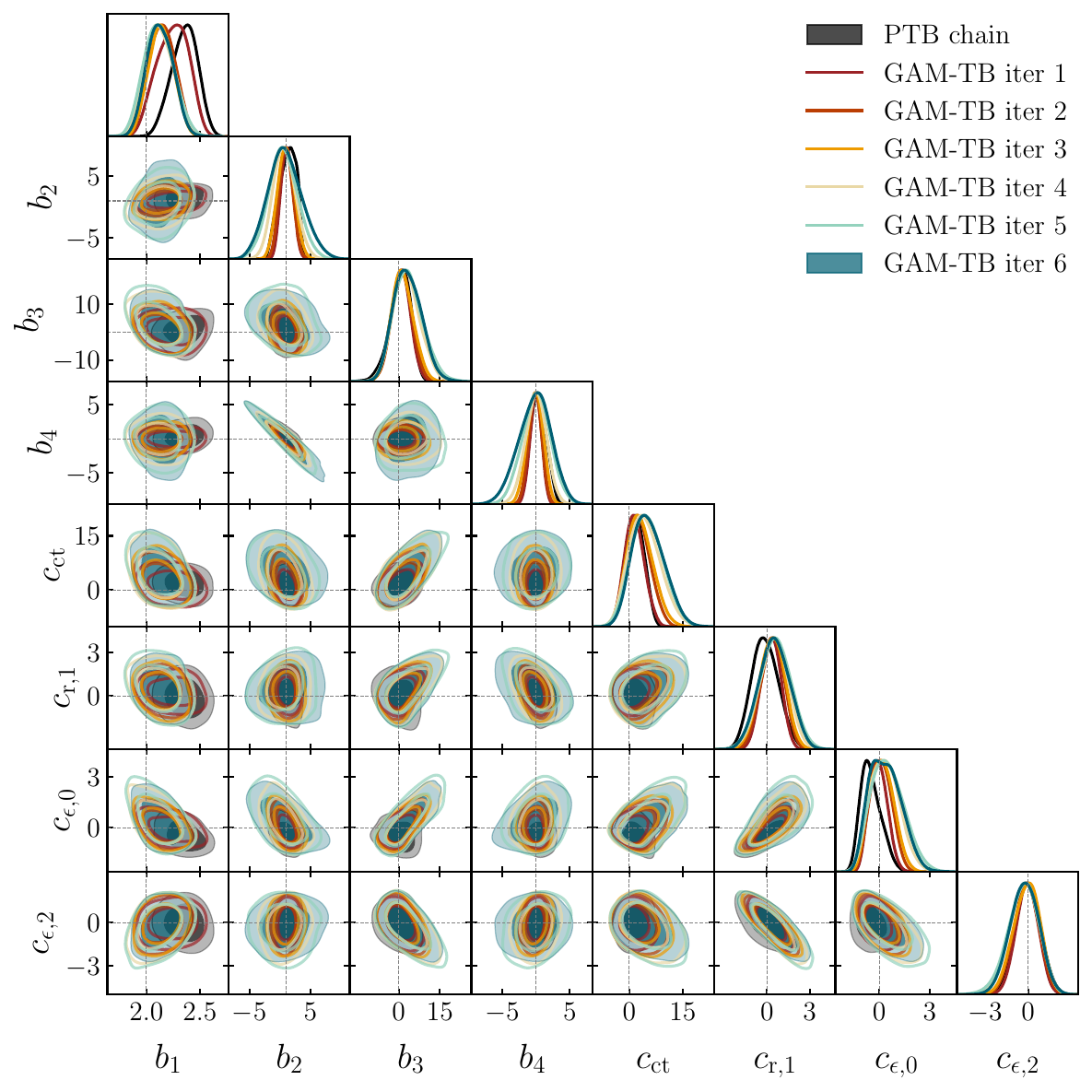}
    \caption{Analogous of Fig.~\ref{fig:LCDM_iter_triangle} but showing the EFT parameters in the standard basis (right) and the 2D plots of cosmological and EFT parameters (left). For the MCMC runs in the rotated basis, we performed the inverse transformation to derive the posterior in the original EFT basis.}
    \label{fig:LCDM_iter_full}
\end{sidewaysfigure}

\subsubsection*{Iterative approach results}

We report here results from the iterative GAM transformation approach described in Sect.~\ref{sec:dataset}. Without any preference, we selected a single dataset among the 100 generated, and iteratively applied the GAM procedure up to six times. Fig.~\ref{fig:LCDM_iter} shows the cosmological parameters estimates (mean and 68\%CI) as function of the iteration, and Figures~\ref{fig:LCDM_iter_triangle} and \ref{fig:LCDM_iter_full} illustrate the full triangle plot for the same chains. More specifically, we report in Fig.~\ref{fig:LCDM_iter_triangle} the triangular block corresponding to the cosmological parameters, and in Fig.~\ref{fig:LCDM_iter_full} the the triangle plot for the EFT parameters and the 2D contour plots from the cosmological parameters-EFT parameters block.

\begin{table}[htbp!]
    \footnotesize
    \centering
    {\renewcommand{\arraystretch}{1.2}
    \begin{tabular}{l|lcccccr}
    \hline
    & \multicolumn{6}{|c}{$\Delta \:[(\theta - \theta_\mathrm{true})/\sigma]$ } \\
    Parameter  & PTB & iter 1 & iter 2 & iter 3 & iter 4 & iter 5 & iter 6 \\
    \hline
    \hline
    $\ln 10^{10}A_{\rm s}$  & \bf -2.3 & \bf -1.2 & \bf -0.8 & \bf-0.8  & \bf -0.6 & \bf -0.7 & \bf-0.7  \\
    $n_{\rm s}$             & -0.13    & -0.004   & -0.15    & -0.18    & -0.14    & -0.18    & -0.4     \\
    $h  $                   & -0.13    & -0.8     & -0.9     & -0.9     & -0.8     & -0.9     & -0.7     \\
    $\omega_{\rm b}$        & 0.0      & 0.0      & 0.0      & -0.0     & -0.0     & -0.0     & -0.0     \\ 
    $\omega_{\rm c}$        & -0.13    & -0.9     & -1.0     & -1.0     & -0.9     & -0.9     & -0.8     \\
    $w_0$                   & \bf -1.0 & \bf -0.9 & \bf -0.6 & \bf -0.7 & \bf -0.7 & \bf -0.7 & \bf -0.7 \\
    \hline
    \end{tabular}}
    \caption{We report the distance from the true input value of the posterior distribution mean in units of $\sigma$, defined as $Delta \:[(\theta - \theta_\mathrm{true})/\sigma]$, for each iteration of the GAM iterative approach to the single dataset as in Fig.~\ref{fig:LCDM_iter} and Fig.~\ref{fig:LCDM_iter_triangle}. We highlighted with bold font the values for $\ln 10^{10}A_{\rm s}$ and $w_0$, exhibiting a $\gtrsim1\sigma$ bias in the PTB.}
    \label{tab:margestats_iterative}
\end{table}

The iterative approach significantly reduces projection effects removing much of the residual correlation seen after a single NL transformation application. This is particularly evident from the strongly degenerate parameters $\ln{10^{10}A_\mathrm{s}}-b_1$: the 1D marginal distribution are closer to the true value for both the parameters, and the 2D contours rotate to align to the axes as the we go through the iterative process. We report in Tab.~\ref{tab:margestats_iterative} the separation of the mean posterior values from the true input cosmological parameters, in units of the standard deviation, for each iteration. We highlight how the iterative application of the GAM transformation allows, in this configuration, to effectively reduce the parameter biases below $ 1\sigma$, especially for $\ln 10^{10}A_{\rm s}$ and $w_0$. On the other hand, we observe larger bias for $h  $ and $\omega_{\rm c}$, possibly due to the strong physical degeneracy between the two cosmological parameters, which is not addressed by the presented methodology. We point out that these results correspond to a single dataset, and  small deviations can arise due to random noise--yet the results remain  compatible with the input cosmological parameters.

As a final consideration, we stopped the iterative process at iteration 6, but cosmological parameter posteriors converged to a common distribution already after three-four iterations, as illustrated in Fig.~\ref{fig:LCDM_iter_triangle} and Tab.~\ref{tab:margestats_iterative}. However, this might not be the case with more complicated cosmological models and larger datasets, and therefore additional iterations might be required to find stable cosmological parameter estimates. The computational time required to achieve convergence increases through iterations--it took $\sim 1h$ runtime for the first iteration and a few hours already for iteration 4--and alternative sampling techniques will help at improving this computational aspect.

\subsection{Is the proposed re-parameterization injecting information?}
\label{subsec:GAM-iteration}

We are using the data itself to determine the prior on the EFT parameters in the re-parameterized space. Consequently, one concern might be that we are somehow injecting additional information through the proposed re-parameterization. In Appendix~\ref{app:prior}  we  consider  a constant likelihood in the TB along with Standard Normal priors on the post-transform nuisance parameters, $N'$, and reverse-transform it back into the PTB. We find that the recovered posterior in the PTB is far wider than the constraints, suggesting that little information is being added. 

A second concern might be that because the proposed approach uses the same data to fit the cosmological model twice (Steps 1 \& 3) that it may under-estimate the true uncertainty. If so, we might expect that iteratively applying the proposed approach to the same dataset would continually reduce the uncertainty. To test this behaviour, we first performed the following test on one of the simulated datasets generated for this work:
\begin{enumerate}
    \item[(i)] Fit the preliminary cosmological model, yielding posterior samples of $\{\mathbf{C},\mathbf{N}\}$.
    \item[(ii)] Fit the orthogonalization model to the preliminary samples of $\{\mathbf{C},\mathbf{N}\}$, to obtain transformation $\{\mathbf{C},\mathbf{N}'\}=\{\mathbf{C},g(\mathbf{C},\mathbf{N})\}$.
    \item[(iii)] Fit the re-parameterized cosmological model using, obtaining posterior samples for $\mathbf{C},\mathbf{N}'$.
    \item[(iv)] Apply the inverse transformation $\{\mathbf{C},\mathbf{N}\}=\{\mathbf{C},g^{-1}(\mathbf{C},\mathbf{N}')\}$ to the posterior samples of $\mathbf{C},\mathbf{N}'$, yielding posterior samples of original (un-rotated) parameters $\mathbf{C},\mathbf{N}$. 
    \item[(v)] Treating these posterior samples of $\{\mathbf{C},\mathbf{N}\}$ as results of a preliminary fit,  repeat steps (ii)--(iv).
\end{enumerate}

\begin{figure}[htbp!]
    \centering
    \includegraphics[width=0.95\linewidth]{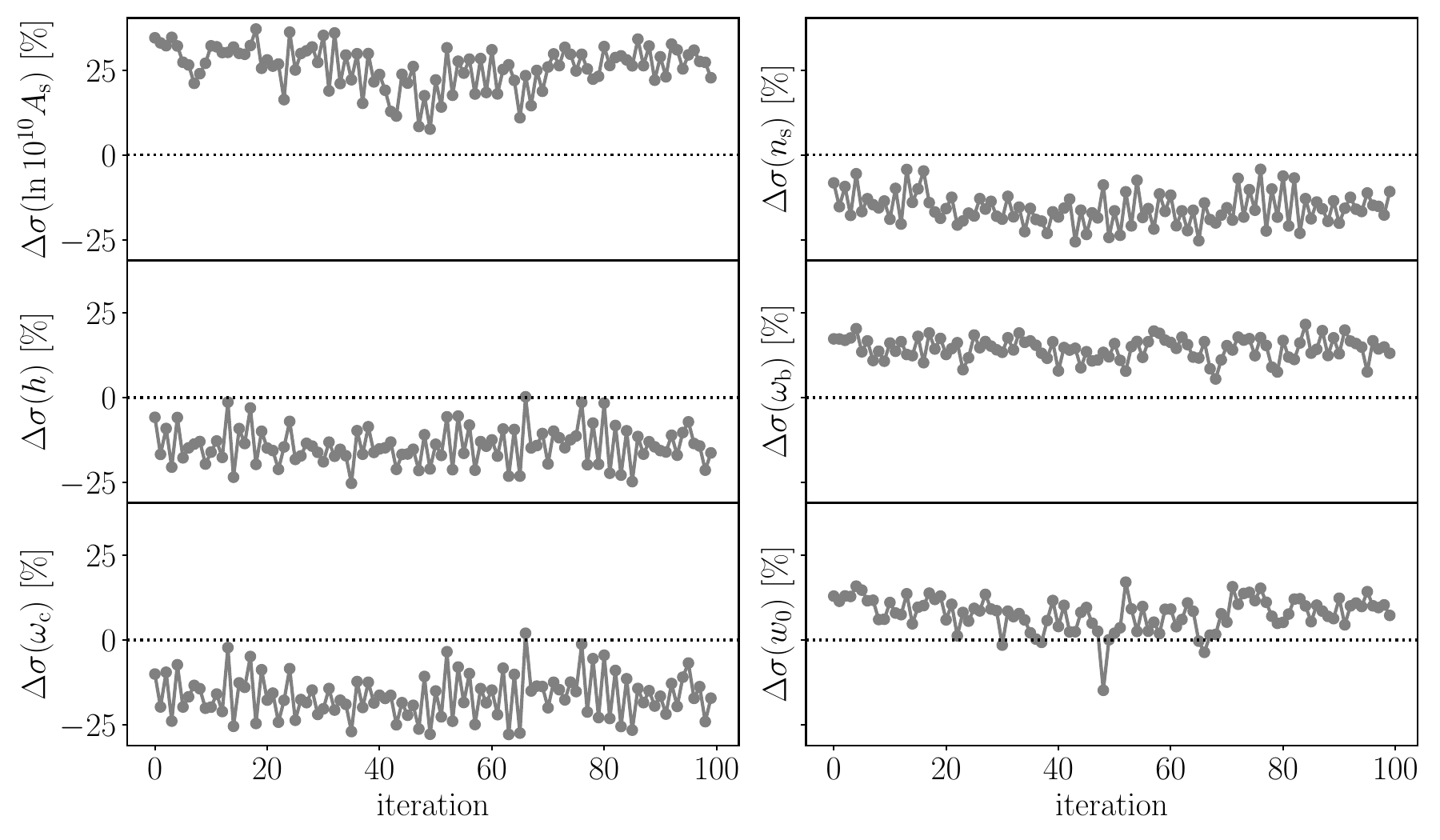}
    \caption{95\% CI range for 100 iterations of the cosmological parameters. Each iteration has been computed after performing a GAM-based transformation on the MCMC samples in the cosmology+EFT space (after transforming back from the GAM-TB) at the previous iteration. The 95\% CI range is expressed in \% of the same quantity from the PTB MCMC run.}
    \label{fig:GAM-iterations}
\end{figure}

The results of this test are summarized in Fig.~\ref{fig:GAM-iterations}. The plot shows the 95\% credible interval widths for the cosmological parameters after each iteration, in units of $\sigma$ from initial preliminary fit (i). Over the course of 100 iterations, we do not observe a decrease in interval width, indicating that the proposed methodology does not continually reduce uncertainty on the cosmological parameters; on the contrary, we find that parameters' uncertainties lie between $\pm25\%$ of the PTB ones.

Even though we do not observe a continual decrease in uncertainty upon applying the method repeatedly, there could still be reduced uncertainty after transformation relative to the PTB. To investigate this, we compute, for each parameter, the ratio $r=\langle\sigma(\theta)\rangle / \sigma(\langle\theta\rangle)$ where the numerator $\langle\sigma(\theta)\rangle$ is the average (across the 100 datasets) of the posterior standard deviations,  and the denominator $\sigma(\langle\theta\rangle)$ is the standard deviation of the marginal posterior means (i.e., the empirical standard error). In a frequentist sense, valid standard errors should yield $r=1.0$, whereas values below 1.0 would indicate  underestimation of the sampling variability in posterior mean estimates. Results are summarized in Table~\ref{tab:std_ratios}. We first find no evidence that the actual sampling variability in estimates is being substantially underestimated: values of $\sigma(\langle\theta\rangle)$ for the GAM are largely consistent with those in the TB. Second we find that the ratios $r$ remain at least 1.0 (and in fact are closer to 1.0 than their PTB counterparts, indicating less overestimation of sampling variability).
A striking value of $r$ in Table~\ref{tab:std_ratios} is the one for $\omega_{\rm b}$, but this does not come as a surprise, because we imposed a strong Gaussian prior on this parameter whose effect is to anchor the mean of the posterior distribution throughout the different realizations, yielding to a value of the empirical standard error much smaller than the average of the posterior standard deviations --the latter reflects the prior's width.

\begin{table}[htbp!]
    \footnotesize
    \centering
    {\renewcommand{\arraystretch}{1.3}
    \begin{tabular}{l|lcr|lcr|lcr}
    \hline
    & \multicolumn{3}{|c|}{$\langle\sigma(\theta)\rangle$} & \multicolumn{3}{|c|}{$\sigma(\langle\theta\rangle)$} &\multicolumn{3}{|c}{$r = \langle\sigma(\theta)\rangle / \sigma(\langle\theta\rangle)$} \\
    Parameter & PTB & LM & GAM & TB & LM & GAM & PTB & LM & GAM \\
    \hline
    \hline
    $\ln 10^{10}A_{\rm s}$  & 0.14 & 0.18 & 0.18 & 0.052 & 0.068 & 0.069 & \bf{2.6} & \bf{2.6} & \bf{2.6}  \\
    $n_{\rm s}$             & 0.061 & 0.052 & 0.057 & 0.040 & 0.041 & 0.042 & \bf{1.5} & \bf{1.2} & \bf{1.3}  \\
    $h$                     & 0.012 & 0.010 & 0.010 & 0.010 & 0.010 & 0.010 & \bf{1.2} & \bf{1.0} & \bf{1.0}  \\
    $\omega_{\rm b}$\tablefootnote{We imposed the Gaussian BBN prior, responsible for this strikingly large value of $r$.} & 0.00030 & 0.00037 & 0.00037 & $4\cdot10^{-5}$ & $1.3\cdot10^{-5}$ & $2\cdot10^{-5}$ & \bf{8.6} & \bf{27} & \bf{18}  \\
    $\omega_{\rm c}$        & 0.009 & 0.007 & 0.008 & 0.007 & 0.008 & 0.008 & \bf{1.3} & \bf{1.0} & \bf{1.0}  \\
    $w_0$                   & 0.25 & 0.28 & 0.28 & 0.16 & 0.16 & 0.16 & \bf{1.6} & \bf{1.8} & \bf{1.8}  \\
    \hline
    \end{tabular}}
    \caption{The average (across the 100 datasets) of the posterior standard deviations, $\langle\sigma(\theta)\rangle$;  the standard deviation of the marginal posterior means, $\sigma(\langle\theta\rangle)$; and their ratio $r = \langle\sigma(\theta)\rangle / \sigma(\langle\theta\rangle)$.}
    \label{tab:std_ratios}
\end{table}

\section{Conclusions}
\label{sec:Conclusions}

We have presented a novel reparameterization approach that reduces dependence on nuisance parameter priors in the Bayesian inference of Effective Field Theory (EFT) models for large-scale structure (LSS) analysis. By applying transformations using either Generalized Additive Models (GAM) or linear-model (LM) orthogonalization techniques, we effectively decouple nuisance parameters from cosmological parameters. This allows us to easily apply separable priors that address the inherent prior-driven projection effects that typically complicate inference in Bayesian cosmology. 

Our approach leverages GAM transformations for their flexibility in capturing non-linear relationships and LM transformations for their efficiency in linear parameter spaces. The GAM transform in particular allows us to achieve a robust reparameterization, enabling clearer interpretation of cosmological posteriors. Using simulations, we have demonstrated that both methods reduce parameter degeneracies effectively, producing marginal distributions that are less sensitive to the choice of prior on nuisance terms. Iterative tests confirm that these transformations do not inject external information, ensuring unbiased parameter estimation across both the transformed and untransformed bases. Finally, we showed that while a single time application of this methodology helps to reduce biases in cosmological parameter estimates, with a few iterations of this procedure we can significantly improve the cosmological parameter estimation robustness reducing biases below the $1\sigma$ level.

While our results show strong gains in decorrelating nuisance and cosmological parameters, certain limitations persist. Particularly, due to the complex, potentially non-additive dependence structure, the GAM leaves some residual coupling remaining after reparametrization. In effect, our transform is limited both by the functional form assumed (e.g., main effects linear model, non-linear additive model) and the space over which it is defined---i.e. the first exploration of the posterior in the PTB. To allow for this, we sampled the PTB space with Standard Normal priors and rescaled the rotated nuisance parameters according to the PTB standard prior amplitude projected onto the TB; in a more general case, we recommend limiting to prior choices in the transformed basis that maintain consistency with the original parameter space, such that the transform covers the volume of interest.

If we chose to apply uninformative Jeffreys priors to all parameters, then we expect that the proposed reparametrization would be unnecessary, as this prior is invariant to parameter transformation.  In contrast, we have introduced a method to rotate into a basis where separable priors can be placed on two sets of parameters to reduce the impact of the one set's priors on the other. A key advantage of our approach, however, is that it allows for informative priors where appropriate---e.g. on cosmological parameters---while still reducing sensitivity to choice of nuisance priors. Similar to the Jeffreys prior, it violates the strong likelihood principle by supporting the definition of priors that depend on the likelihood.

To apply this method to real-world data, such as the recent DESI DR1 Full Shape measurements~\cite{DESI2024-fullshape, DESI:2025fxa}, several enhancements are necessary to handle the increased complexity of practical analyses. The primary challenge is computational cost: the DESI FS measurements span six different redshift bins, resulting in a much larger number of nuisance parameters. While the analyses presented in this paper achieved convergence within a few hours for the final GAM iteration, it is likely that a DESI-like analysis would require more than a week for the chains to converge. Additionally, given the increased complexity of the posterior surface, it is possible that a higher number of GAM iterations are needed in order to remove the degeneracies and, hence, the projection effects.
Therefore, a key area for improvement is the development of a differentiable GAM. Utilizing a differentiable GAM would allow us to leverage advanced samplers such as Hamiltonian Monte Carlo (HMC)~\cite{betancourt}, which offer superior convergence properties, particularly in high-dimensional parameter spaces, as highlighted from recent applications to the analysis of cosmological summary statistics~\citep{Campagne:2023ter, Piras:2023aub, Bonici:2023xjk, Ruiz-Zapatero:2023hdf, Cagliari:2023mkq, Balkenhol:2024sbv, Mootoovaloo:2024lpv, Giovanetti:2024zce, SPT-3G:2024atg}. Once a differentiable GAM becomes available, \texttt{Effort.jl}’s inherent differentiability and its integration with the probabilistic programming language \texttt{Turing.jl}~\cite{turing} will automatically enable the use of HMC sampling.

A second avenue for development involves more sophisticated reparameterization strategies capable of handling multiple parameters simultaneously~\cite{DGAM, 2024arXiv240720995V}. While the current GAM framework models one parameter at a time, recent advancements have introduced generalizations that allow for simultaneous reparameterization. This could help address residual, unaccounted-for degeneracies and further enhance the robustness of this approach when applied to more complex models.

This reparameterization technique marks an advancement for EFT-based cosmological analyses by enabling more reliable and interpretable parameter estimates. The methodology significantly reduces the dependence of cosmological inferences on nuisance parameter priors, thus mitigating one of the major challenges in Bayesian inference within the EFT framework.

\acknowledgments

SP, MB, MC, WP and GM are supported in part by funding from the Government of Canada’s New Frontiers in Research Fund (NFRF). WP and MB also acknowledge the support of the Canadian Space Agency and the Natural Sciences and Engineering Research Council of Canada (NSERC), [funding reference number RGPIN-2019-03908]. GM acknowledges the support of the Natural Sciences and Engineering Research Council
of Canada (NSERC), [RGPIN-2022-03068 and DGECR-2022-004433].

Research at Perimeter Institute is supported in part by the Government of Canada through the Department of Innovation, Science and Economic Development Canada and by the Province of Ontario through the Ministry of Colleges and Universities. 

This research was enabled in part by support provided by Compute Ontario (computeontario.ca) and the Digital Research Alliance of Canada (alliancecan.ca).

\appendix

\section{GAM prediction check.}
\label{app:GAM_predcheck}

\begin{figure}
    \centering
    \includegraphics[width=0.95\linewidth]{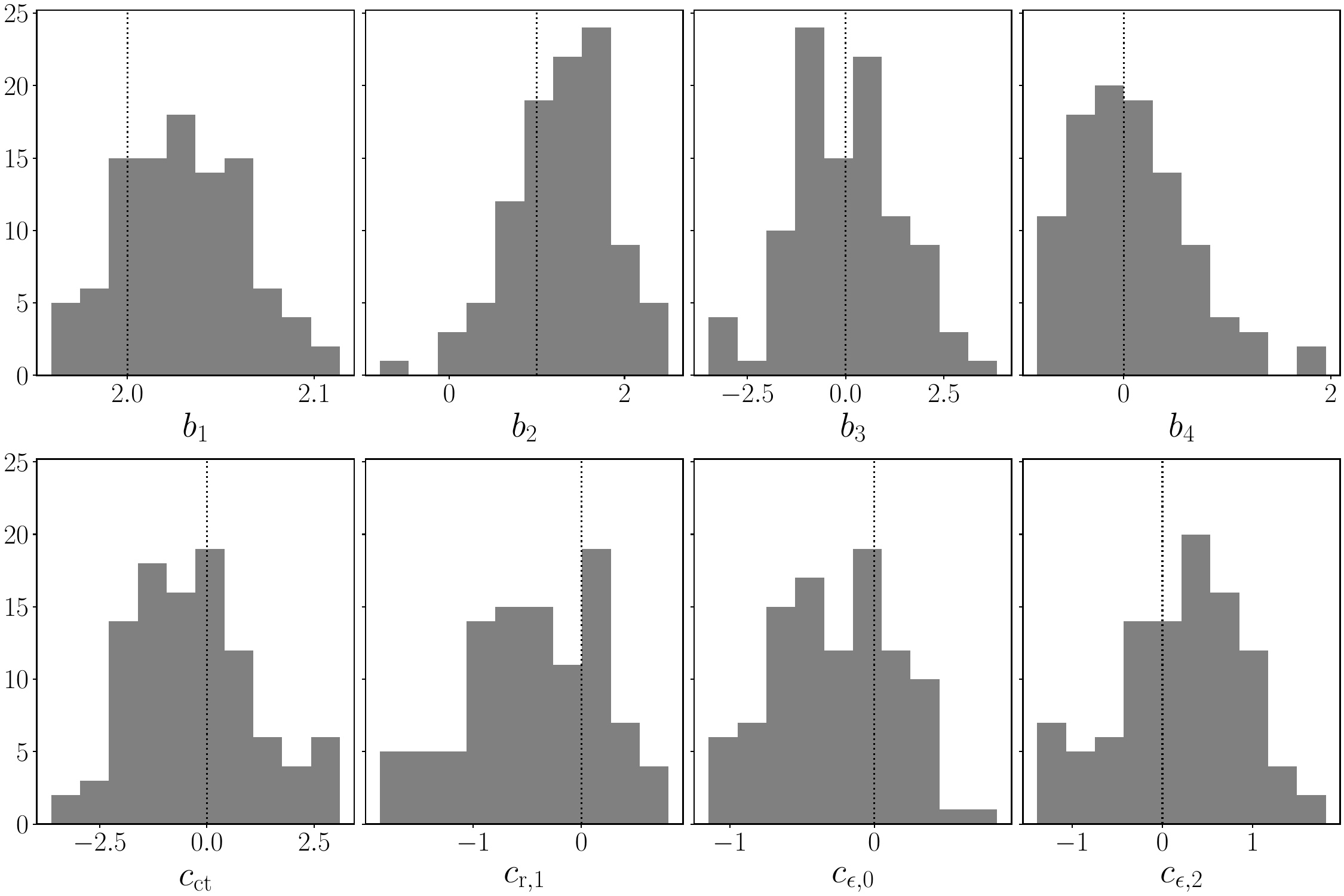}
    \caption{Distribution of EFT parameters predicted by the GAM fits from 100 simulations sharing the same input cosmology but different noise realizations.}
    \label{fig:GAM_prediction}
\end{figure}

A question one might ask is whether the GAM transformation, despite being tuned on MCMC samples that shows projection effects on cosmological parameters, correctly maps the input cosmological parameters into the input EFT parameters. To test this, we have undertaken the following:
\begin{enumerate}
    \item We generated 100 datasets (monopole and quadrupole galaxy power spectra, as in the main analysis) all sharing the same cosmological and EFT nuisance parameter values, but differing on the noise, which was drawn from the NCVM described in Sect.~\ref{sec:dataset}.
    \item We evaluated the GAM transformation for each of the 100 realizations.
    \item We computed the predicted EFT parameters in the un-rotated basis according to Eq.~\ref{eq:GAM_transform} for each of the 100 GAM transformations, by plugging in the true cosmological input parameters.
\end{enumerate}
The GAM transformation are themselves a random variable in this context, likewise the EFT parameters prediction generated in such a way. We therefore show the distribution of the EFT parameters computed via the 100 GAM transformations assuming the true input cosmology in Fig.~\ref{fig:GAM_prediction}. We find that the GAM transformation correctly maps the true input cosmology onto the correct EFT parameters with high confidence, despite each transformation being computed from a posterior that gives biased cosmological parameters' results.

\section{Induced prior on EFTofLSS parameters after GAM parameter space transformation.}
\label{app:prior}

\begin{figure}
    \centering
    \includegraphics[width=0.95\linewidth]{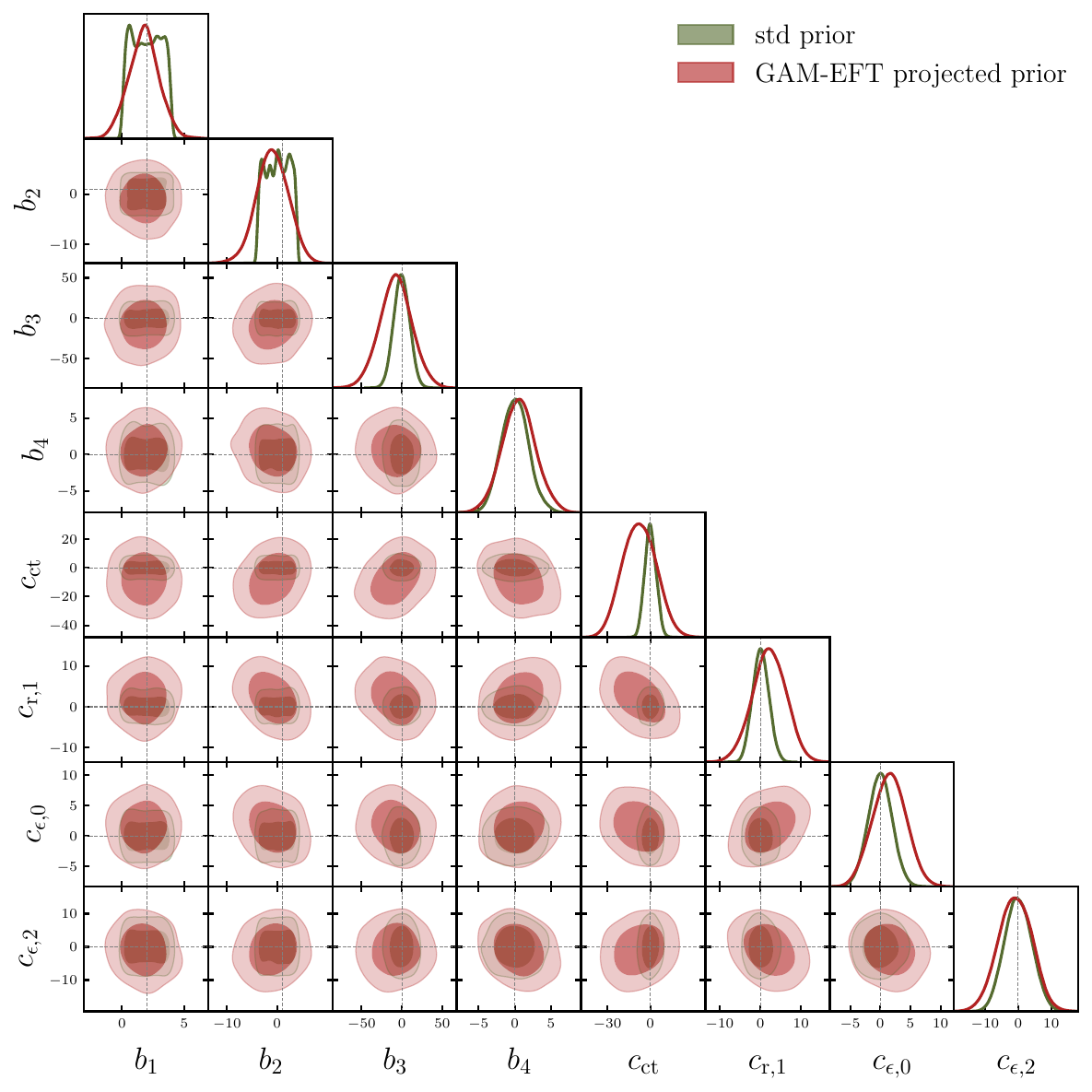}
    \caption{The red contours shows the induced GAM-TB priors on the EFTofLSS parameters. We report the standard EFT priors (green) for comparison, and true input parameters' values as vertical lines.}
    \label{fig:flatGAM-induced_priors}
\end{figure}

In order to show the impact of the transform as discussed in Section~\ref{sec:Results} on the nuisance parameters, we consider what the Standard Normal priors on the nuisance parameters in the TB looks like in the PTB, for a single iteration of the proposed methodology. To do this, we  perform a simple MCMC run with a constant likelihood and uniform priors in the GAM-TB. We show in Fig.~\ref{fig:flatGAM-induced_priors} the induced prior in the un-rotated basis. In particular, we point out that the PTB EFT induced priors are always comparable to the standard priors in the PTB  (green).

\bibliographystyle{JHEP}
\bibliography{biblio.bib}

\end{document}